\documentclass[preprint]{revtex4-2}

\usepackage{graphicx}
\usepackage{tabularx}
\usepackage{dcolumn}
\usepackage{bm}
\usepackage{units}

\usepackage{xcolor}

\begin{document}

\title{Drag on a partially immersed sphere at the capillary scale}
\author{Robert Hunt$^{1}$}
\author{Ze Zhao$^{2}$}
\author{Eli Silver$^{1}$}
\author{Jinhui Yan$^{2}$}
\author{Yuri Bazilevs$^{1}$}
\author{Daniel M. Harris$^{1}$}
 \email{daniel\_harris3@brown.edu}
\affiliation{%
$^{1}$School of Engineering, Brown University\\
$^{2}$The Grainger College of Engineering, University of Illinois Urbana-Champagne
}%

\date{\today}
\begin{abstract}
We study the drag on a centimetric sphere in a uniform flow in the presence of a free surface as a function of submergence depth.  Through direct force measurements in a custom benchtop recirculating flume, we demonstrate that the drag can significantly exceed the corresponding drag in a single-phase flow and achieves a peak at submergence depths just prior to complete immersion.  The additional drag in the partially immersed state is rationalized by considering hydrostatic effects associated with the asymmetric surface height profile induced by the obstacle in the flow direction which persists for flow speeds below the minimum capillary-gravity wave speed.  At these scales, the sphere's wettability plays a pronounced role in determining the maximum possible drag and results in hysteretic behaviors near touchdown and complete immersion.  The influence of flow speed, sphere size, and surface tension on the drag characteristics are additionally explored through a combination of experiments and numerical simulations.

\end{abstract}

\maketitle

\section{\label{sec:level1} Introduction}

Drag on a solid sphere in a single-phase flow is perhaps one of the most fundamental, broadly applicable, and thoroughly studied problems in fluid dynamics. For the case of a Newtonian fluid, the drag on a smooth sphere can be characterized by a single
characteristic curve for the coefficient of drag $C_D$ ($C_D =
\frac{F_D}{\frac{1}{2} \rho U^2 \pi \left( \frac{d}{2} \right)^2}$) as a
function of Reynolds number $Re$ ($Re = \frac{\rho U d}{\mu}$),
where $\rho$ is the fluid density, $U$ is the free stream velocity, $d$ is the
sphere diameter, and $\mu$ is the dynamic viscosity of the fluid.  Less attention has been given to a sphere that is
partially immersed at a fluid-fluid interface, although it is well known that new drag mechanisms are possible near free surfaces due to the additional physics involved. Attempts at extrapolation from the single-phase data are thus bound to underpredict the actual drag on a sphere moving along an interface. 

John Scott Russell performed seminal experiments measuring the drag on partially immersed objects by towing ships in a canal with a varying force and measuring their speed \cite{russell1839iii}. Reech and Froude continued this work \cite{reech1852course, froude1887},
characterizing the new physics at the free surface by a scaling later referred to as the Froude number (or Reech-Froude number)
$Fr$ ($Fr = \frac{U}{\sqrt{g d}}$), where $g$ is the acceleration due to gravity.  Motion at or near a free surface is often associated with an additional drag mechanism commonly referred to as wave drag.  Even in a purely inviscid flow, obstacles create a
moving disturbance of the free surface and generate waves that carry
momentum away from the object.  This radiated energy of course comes at a cost, resulting in a reciprocated non-zero drag force on the solid body. Lord Kelvin \cite{thomson1887} was the first to introduce a theoretical model to calculate the wave resistance using these ideas.  In the same context, Lamb
calculated the wave drag in a 2D flow on a submerged cylinder \cite{lamb1913} using potential flow theory, which Havelock later extended for a submerged sphere \cite{havelock1917some}.

Naturally, most of the first studies of wave drag are in the context of
ship design \cite{froude1887,newman1977marine}.  Other recent studies in the area have focused on submarines \cite{dawson2014investigation}, autonomous marine vehicles \cite{spino2022development}, human swimmers \cite{vennell2006wave, james2016search}, and surface dwelling insects \cite{jami2021overcoming}, to name a few.

Surprisingly, there exists only limited experimental work on geometrically simple objects moving along an interface.  In 1947, Hay considered the drag on towed, vertically oriented, surface-piercing cylinders as a function of diameter, speed, and submergence depth \cite{hay1947}, which was later expanded upon by Chaplin and Teigan \cite{chaplin2003steady}.  While a drag increase over the expected single-phase drag (i.e. in the absence of a free surface) was measured, Chaplin and Teigan accounted for the measured increase in their experiments with a different physical mechanism: the free surface rises on the upstream face of the cylinder, resulting in an additional pressure forcing on the cylinder front that can be approximated from hydrostatics.  The proposed hydrostatic pressure contribution in the upstream region above the quiescent water line was confirmed by direct pressure measurements along the cylinder length for low and moderate $Fr$.  The proposed scaling and experimental results both suggested a value of $Fr$ at which this additional contribution to the drag coefficient should be maximal, with the numerical value depending on the aspect ratio of the cylinder.  They were thus able to rationalize the increased drag observed at an interface without appealing to the traditional arguments associated with excitation of propagating water waves.

When considering the effect of surface tension on the free surface dynamics, travelling waves are predicted to exist only above a minimum wave phase speed. For instance, for the case of a clean air-water interface, this value is 23 cm/s.

In 1996, Raphaël and DeGennes discussed the capillary-gravity wave drag of a
surface point disturbance \cite{raphael1996} and predicted exactly zero wave drag when an object moves slower than this minimum wave speed.

Browaeys et al. \cite{browaeys2001capillary} measured the force on a fixed vertical wire immersed in a rotating circular channel and observed a sharp increase in drag, consistent with the theoretical prediction of Raphaël and DeGennes \cite{raphael1996}. 
Burghelea and Steinberg \cite{burghelea2001,burghelea2002} performed a similar experiment with a fixed, millimetric sphere and observe a continuous transition, attributing the discontinuity observed in Browaeys et al. \cite{browaeys2001capillary} to experimental noise in the region of transition.
Chevy and Raphaël \cite{chevy2003capillary} present a theoretical model which includes the effect of force variation due the imposition of fixed depth of the travelling surface disturbance, which causes the wave resistance to vary continuously across the critical wave speed, in qualitative agreement with Burghelea and Steinberg \cite{burghelea2001,burghelea2002}. 
Richard and Raphaël \cite{richard1999capillary} introduce viscosity to the model of Raphaël and DeGennes \cite{raphael1996} and show that, unlike the inviscid theory, the wave resistance remains bounded near the critical wave speed, and a small contribution to the resistance emerges below the critical wave speed.  Most recently, Merrer et al. \cite{merrer2011} performed experiments measuring the deceleration of liquid nitrogen droplets sliding along a liquid bath, which showed good agreement with the classical wave drag theories.

For motion along an interface at small scales, narrowing focus solely on momentum transfer to surface waves has led to prior paradoxes such as Denny's paradox for water striders \cite{denny1993air}, 
which was eventually resolved by appealing to other complementary aspects of the underlying bulk fluid physics \cite{hu2003hydrodynamics}.

Due to its geometric simplicity, in the present work we focus on the case of a partially immersed solid sphere in a steady uniform flow.  Only a few other works have focused on this configuration, the most relevant of which we review in what follows.

Burghelea and Steinberg \cite{burghelea2001,burghelea2002} focused on the onset of wave resistance when traversing the minimum capillary-gravity wave speed with a vertically suspended, surface-piercing millimetric sphere. Only a single submergence depth was considered.
They fit a second-order polynomial to the measured drag below the transition and use the difference in force from this curve, extrapolated above the critical wave speed, to deduce that the transition to wave drag is continuous.

Benusiglio et al. \cite{benusiglio2015wave} measured the drag on fully submerged centimetric spheres connected to a torsion balance with a second, deeply submerged sphere, which allows for direct measurement of the net drag difference due to the free surface. They observe a maximum of this difference as a function of speed and show that the additional free surface drag is on the order of the fully submerged drag. This measured drag difference is compared with Havelock's inviscid gravity wave model and shown to be an order of magnitude smaller than predicted, with the difference attributed to a saturation of the generated wave amplitude.

Kamoliddinov et al. \cite{kamoliddinov2021} consider a floating 10 cm sphere with density chosen so that the static sphere floats at half submergence. A fishing line is attached to the sphere and used with a weight and pulley system to tow the sphere horizontally with a constant force. They observe horizontal oscillations for $Fr < 0.6$, a significant free surface deformation and gradual descent of the sphere position for $0.6 < Fr < 1.2$, and vertical oscillations for $Fr>1.2$.  Although the vertical and lateral positions of the sphere are not constrained, an increased average drag coefficient (above that expected by considering half of a fully immersed sphere) is observed for all speeds, with a peak near $Fr\approx1$.

Of perhaps most direct similarity to the current study, James et al. \cite{james2015,james2016search} studied rigidly constrained, partially submerged decimetric spheres which are towed across a range of speeds for four submergence depths. They observe a force maximum as a function of $Fr$ for each submergence depth and propose a mechanism for this critical $Fr$, whose value depends on submergence depth in their study.  Their model estimates the run-up height on the upstream face of the sphere and correlates the critical $Fr$ with the run-up height needed to overcome the top of the sphere in a so-called $Fr$ transition.  These measurements are performed at significantly larger flow speeds, length scales, and Reynolds numbers than the present work.

In this work, we present detailed direct force measurements of the drag on a fixed centimetric sphere as a function of submergence depth.  For this purpose, a custom small-scale benchtop recirculating water flume is developed and characterized.  As the sphere is lowered into the flow, the drag steadily increases with submergence depth.  As it nears complete submergence, we observe a significant drag enhancement over the fully immersed single-phase scenario as a result of the deformations of the interface.  The additional drag associated with the free surface in the partially immersed state is well approximated by considering the contribution of additional unbalanced hydrostatic forces stemming from the asymmetric surface deformation, in a similar mechanism to that described by Chaplin and Teigan for the surface piercing cylinder \cite{chaplin2003steady}.  While the flow speed and sphere radius have only a modest impact on the drag coefficient $C_D$ for the regime considered, we find that the sphere wettability plays a significant role in determining the drag and results in new hysteretic behavior.  Furthermore, the influence of surface tension on the documented force curves is explored in a series of companion numerical simulations.

\begin{figure*}
\includegraphics[width=\textwidth]{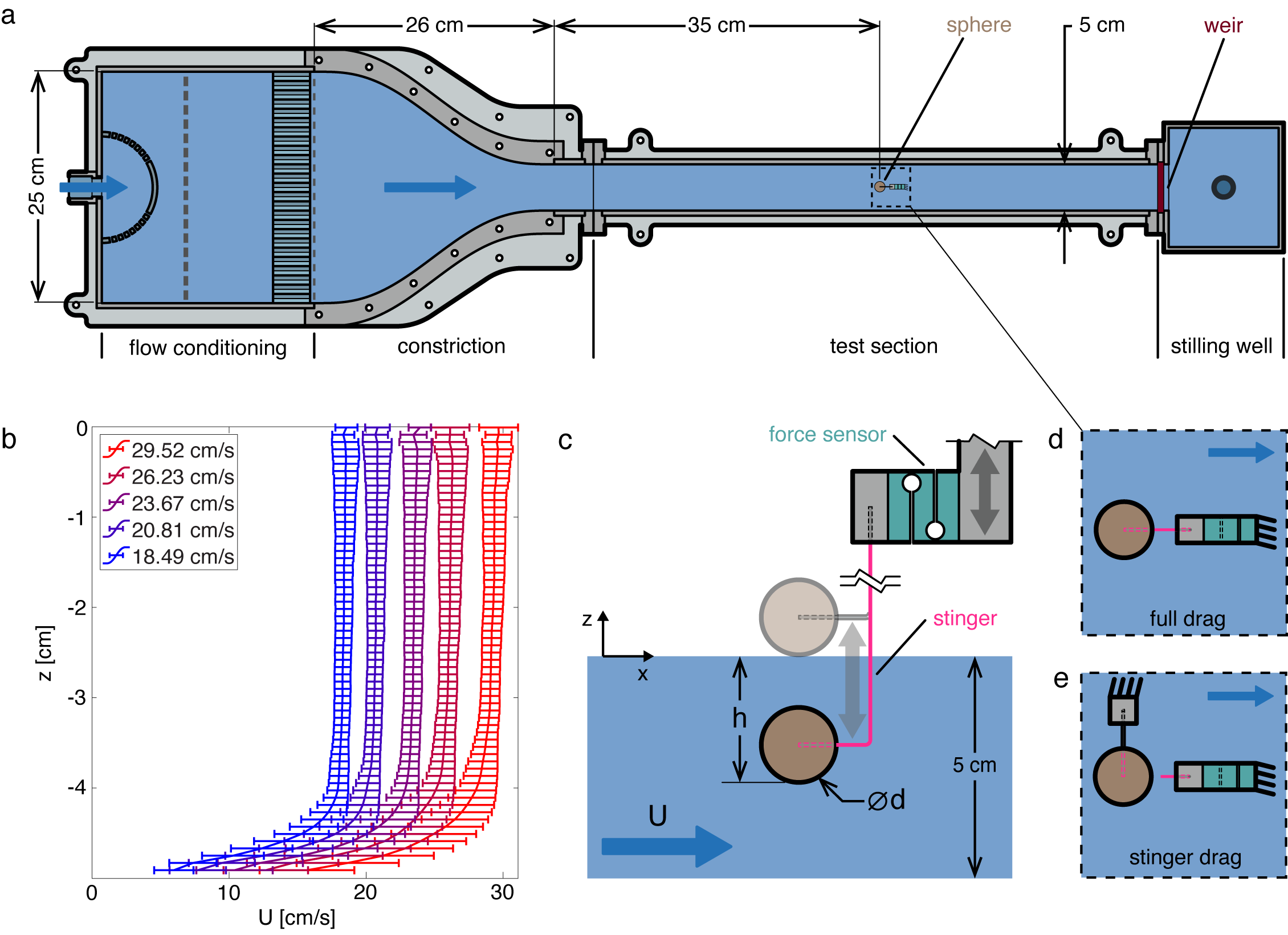}
\caption{\label{Overview}(a) Top view of the custom flume used for drag measurement experiments. (b)
  Experimental measurements of streamise velocity along the center of the
  channel. Error bars indicate one standard deviation of horizontal velocity
  values. Legend entries indicate the nominal velocity value (average value
  over $z \in [- 1, 0]$ cm) used for computation of experimental drag coefficients. (c) Side view of sphere, stinger, and
  force sensor. (d) Top view of standard drag measurement configuration, which
  includes contributions from the sphere and stinger. (e) Top view of alternate
  configuration used to measure forces on the stinger only in the wake of an independently constrained sphere.}
\end{figure*}

\section{\label{ExpMethods} Experimental methods}
To perform direct measurements of the sphere drag, we designed, constructed, and characterized a custom recirculating water
channel consisting of a pump, flow conditioning section, constriction, test section,
weir, and a stilling well shown schematically in Figure \ref{Overview}(a). Drag measurements are performed in the test section
downstream of the constriction. A single sphere is held centered in the
channel by a stinger attached in the downstream wake region of the sphere. The
stinger is rigidly mounted to a force sensor fixed to a motorized linear stage, allowing for
fine control of the sphere's vertical position. Additional details regarding the flume
 and force measurements are discussed in the following sections.  The range of dimensional and non-dimensional parameters explored in the present experimental work are summarized in Table \ref{values}.
 \begin{table*}
 \def\arraystretch{1.15}
\begin{ruledtabular}
\begin{tabular}{cccc}
parameter & symbol & definition & value \\
 \hline\hline
 sphere diameter & $d$ & -- & $7.8-12.7$ mm \\ 
 \hline
 submergence depth & $h$ & -- & -$10-27.5$ mm \\
 \hline
 flow speed & $U$ & -- & $18.49-29.52$ cm s$^{-1}$ \\
 \hline
 density (water) & $\rho$ & -- & $.997$ g cm$^{-3}$ \\
 \hline
  kinematic viscosity (water) & $\nu$ & -- & $.01$ cm$^{2}$ s$^{-1}$  \\
 \hline
  surface tension & $\gamma$ & -- & $72.2$ mN m$^{-1}$  \\
 \hline
  gravitational acceleration & $g$ & -- & $981$ cm s$^{2}$  \\
 \hline
  Reynolds number & $Re$ & ${U d}$ / ${\nu}$ & $2070 - 3750$  \\
 \hline
  Froude number & $Fr$ & ${U}$ / ${\sqrt{g d}}$ & $.52 - .94$  \\
 \hline
    Bond number & $Bo$ & ${\rho g d^{2}}$/${4 \gamma}$ & $2.1- 5.5$  \\
 \hline
   Weber number & $We$ & $\rho U^{2} d$ / $\gamma$ & $6.0 - 15.3$  \\
    \hline
 coefficient of drag &$C_{D}$ & $F_{D} / {\frac{1}{2} \rho U^2 \pi \left( \frac{d}{2} \right)^2}$ & $ - $  \\
 \hline
  coefficient of drag, free stream &$C_{D,0}$ & $F_{D,0} / {\frac{1}{2} \rho U^2 \pi \left( \frac{d}{2} \right)^2}$ & $ 0.41 - 0.43$  \cite{morrison2013introduction}  \\
\end{tabular}
\end{ruledtabular}
\caption{\label{values}Relevant parameter symbols, definitions, and values associated with our experimental study.}
\end{table*}

\subsection{\label{FlumeDesign}Flume design and characterization}
We manufactured a custom recirculating water channel using primarily laser cut
acrylic and 3D-printed parts. A 1/5 hp centrifugal pump (Meziere WPX750) is
used to pump water through a 3D-printed diffuser into a 25 cm wide inlet
stilling area. This allows for the flow to stagnate before passing through an
aluminum honeycomb flow conditioner of length 4.0 cm with approximately 1.0 cm
wide tubules. After passing through the flow conditioner, the water enters a
converging section which reduces the width from 25 cm to the test section
width of 5.0 cm, with a constriction ratio chosen as large as possible given our design constraints \cite{bell1988contraction}. The profile of this section was designed in Rhinocerous 3D software as
a fifth-order polynomial curve enforcing continuity of curvature at the endpoints.
This profile was manufactured using fourteen stacked 5.5 mm thick acrylic sheets
compressed together using stainless steel through bolts. Two 12.7 mm thick
acrylic flanges with machined mating faces and a compression seal are used to
connect the converging section with the test section. Each section is designed
with mounting points corresponding to a 25.4 mm grid to allow for installation
on an optical table (Thorlabs Nexus B3648T) with passive vibration dampers (McMaster-Carr 62075K21) to isolate the flume from external vibrations. The optical table and dampers are
installed on an extruded aluminum frame. The open test section is manufactured from
three 5.5 mm thick acrylic panels with a length of 59 cm and assembled to form a 5.0 cm wide by 8.0 cm tall
interior cross section. At the end of the test section, laser-cut acrylic weir
gates are used to control the water height in the test section. After spilling
over the top of the weir, the water enters a 48 cm deep stilling well. Water
is added to the recirculating flume until the water height in the stilling
well reaches just below the bottom of the weir gate. The water spilling into
the stilling well entrains air, and the dimensions of the stilling well were
chosen to allow entrained bubbles to rise to the free surface without being
recirculated. Two laser-cut diffusers are placed below the surface of the
stilling well to diffuse the jet leaving the test section to further subdue
air entrainment.

To vary the speed in the flume while keeping the water height in the test
section fixed at 5.0 cm, gates of height 18, 20, 22, 24, and 26 mm are used
for the weir, with taller gates corresponding to lower speeds. Once a particular weir is installed, the variable speed setting on the pump controller is adjusted until the water
height in the test section reaches 5.0 cm.

To characterize the flow profile in the flume, particle image velocimetry
(PIV) measurements were performed in the test section. A dual-head Nd:YAG laser
(New Wave Solo-I) enters a custom sheet optic allowing for adjustment of the
sheet thickness and width to approximately 1 mm and 7.0 cm, respectively. This
sheet is reflected through the bottom of the test section to form a vertical plane
aligned with the streamwise direction. PIV image pairs taken
1.0 ms apart are recorded at a rate of 15 Hz using a Allied Vision Mako U-130B
CMOS camera with 1 MP resolution. Signals for the laser and camera trigger are
generated using custom Arduino code. The delay time between pulses is
confirmed independently using a high-speed camera. PIV images are processed
using PIVlab \cite{thielicke2021} in MATLAB using a 32 $\times$ 32 px interrogation
window with 50\% overlap. Images are calibrated using a custom laser-etched
calibration plate. We estimate the error contribution to the measured velocity from the length and time calibration to be less than 1\%. Velocity values along the centerline as a function of depth at the sphere measurement location
are reported in Figure \ref{Overview}(b). These values are aggregated over a 2.4 cm wide
area for 13 seconds. The average horizontal speed taken over the region $h \in
[- 1.0 \ \mathrm{cm}, 0.0 \ \mathrm{cm}]$ is used as a nominal value for $U$ and used
for calculations of the mean drag when reporting $C_D$. The extrema of
velocity over the region $h \in [- 2.75 \ \mathrm{cm}, 0.0 \ \mathrm{cm}]$,
corresponding to the total region traversed by the sphere for force
measurements reported herein, are used in calculating the uncertainty in $U$ when force measurements are
presented as $C_D$. The mean 2D turbulence intensity value taken over the total region traversed by the sphere for all speeds presented here is 3.1\%.

\subsection{\label{Spheres}Sphere manufacturing and surface treatment}

Direct force measurements typically involve a rigid, thin body, referred to as a
stinger, to constrain the forced body.  To install the stinger, acrylic spheres are compressed between two 3D printed drill guides with a thin
layer of nitrile rubber to inhibit movement of the sphere relative to the
drill guide. A drill press is used with a drill bit of 0.6 mm diameter to bore
a hole to the center of the sphere. The stinger, consisting of a .62 mm diameter high strength steel wire, is inserted and held in place
with superglue and bent upwards at a right angle 5 mm away from the
surface of the sphere. This bend is positioned in the downstream wake of the
sphere to avoid interactions with the contact line near the top or bottom of
the sphere as shown in Figure \ref{Overview}(c). 

To manufacture superhydrophobic spheres, two-part NeverWet coating is applied to
 acrylic spheres as in prior work \cite{galeano2021capillary}, and the stinger installed using the method described previously. During the spray coating process, the stinger is masked with Kapton tape to avoid also
coating with NeverWet.

To manufacture superhydrophilic spheres, agar hydrogel (Acros Organics
400405000) is cast in a 3D-printed mold. The mold is prepared with two
sections and two holes, one hole to insert the stinger, and one hole to inject
the liquid agar mixture. Agar powder is weighed, and boiling distilled water
is added and weighed to yield 4\%w/w agar to water. This solution is mixed
using an immersion blender. A syringe is filled with the liquid agar, then a
needle is attached to the syringe. After purging air, the liquid mixture is
injected into the mold and left to solidify. The mold is removed after
approximately 15 minutes. To better adhere the stinger to the cast agar, the
part of the stinger tip which is embedded in the agar sphere is wound into a helical
shape before casting. A very small seam on the cast sphere at the mold junction was designed to be co-planar with the stinger (aligned with the free stream flow direction) to minimize any possible effect on the projected area and contact line. Although cast agar is a soft elastic material, for a value of the Young's modulus $E = 540.2$ kPa as reported previously \cite{EDDAOUI2019746}, a dynamic pressure estimate corresponding to the maximum flow velocity presented here yields a strain of less than .01\%.

The nominal equilibrium contact angles for the agar, acrylic, and NeverWet coated acrylic spheres are $15^{\circ},70^{\circ},160^{\circ},$ respectively as determined from prior direct measurements in the literature \cite{andersson2017dynamics,osti2009,weisensee2017droplet}.

\subsection{\label{ForceSensor}Force measurements}
To conduct force measurements, a Futek Miniature S-Beam Jr. load cell (FSH03867)
force sensor with a maximum load of 10 grams is used. The stinger from the sphere
is connected to the force sensor by a 3D-printed coupler. The force sensor
is connected to the Futek IAA100 strain gauge amplifier. The amplifier gain
is adjusted to increase the sensitivity of the output and calibrated using a
linear fit of 5 test weights independently measured using a magnetic force balance scale (US Solid USS-DBS5) over a range of 24.5 mN. The output from the amplifier is input to an ADS1115 16-bit analog to digital converter connected
to an Elegoo Mega R3 with ATMega2560 microcontroller. Force samples are
collected at a mean sample rate of 11 Hz.

In our testing of many possible stinger configurations, we observed the most significant
significant effect on the drag characteristics when the stinger was attached near the top
or bottom of the sphere.  When the stinger is
connected to the sphere in the downstream wake region (Figure \ref{Overview}(c,d)), there is only minimal effect on the flow and free surface.  To directly
account for the additional drag introduced by the stinger, we measure
the force on the stinger while disconnected from the sphere but held in place
behind a fixed sphere, (Figure \ref{Overview}(e)) scanning the depth in both descending 
and ascending directions.  A similar taring procedure and stinger arrangement was proposed by James \cite{james2016search}.  We fit a third-order polynomial to the measured
stinger drag and subtract it from the full drag obtained for the combined
sphere and stinger to arrive at the drag values for the solitary sphere. The measured stinger drag was smaller than $0.35$ mN for all partially immersed cases.

The force sensor is attached to a linear stage actuator (Befenbay BE069-4)
with 5 mm pitch driven by a NEMA 17 stepper motor and driver (HoCenWay DM556).
The AccelStepper Arduino library is used to control the sphere's vertical
position for the drag measurements. Force readings are acquired continuously
as the sphere moved smoothly from approximately 10 mm above the surface to 35 mm below the surface and back to the initial position over approximately 400 seconds total.  Force readings are binned within 0.25 mm increments of height.  The variation of the mean
force over this distance is significantly smaller than the observed unsteady force fluctuations.

With the exception of the single trial data presented in Figure \ref{Single}, each curve presented here comprises the mean of 5 independent trials. The uncertainty for these reported values is calculated by the estimator of the standard error of the mean, $\frac{\sigma}{\sqrt{N}}$, where $\sigma$ is the corrected standard deviation of the sampled means and $N$ is the number of trials. It is worth noting that these uncertainties do not include the unsteady force fluctuations.  The standard deviation of fluctuations at each height for a typical trial is presented in Figure \ref{Single}.  

To calculate the uncertainty for drag forces reported dimensionally (mN), the
nonlinearity ($\pm .098$ mN), hysteresis ($\pm .098$ mN), and repeatability ($\pm .049$ mN) errors from the force sensor documentation are
added in quadrature with the standard error of the mean. To
include the uncertainty of the flume velocity in the force measurements
reported as coefficient of drag $C_D$, we use extrema of the mean velocity
from PIV taken over the reported submersion depths as worst-case estimators of
error.

\subsection{\label{FreeSurfaceSilhouetteMeasurement}Free-surface measurement}
To measure the silhouette of the free surface from the side, a lens (United
Scientific LCV108) and reflecting mirror is placed across the channel from the
camera, with an LED light panel partially covered by colored diffuser paper
placed at the focal plane of the lens. This allows for the bulk of the fluid
to appear bright, while refracted rays passing through the free surface appear
dark. The meniscus at the side of the channel also appears dark due to this effect. Stepping and camera triggering are controlled directly from the Arduino.
A Nikon D850 with Nikon Micro-NIKKOR 105 mm macro lens is used in conjunction
with a relay module connected as remote shutter controller. In the acquired
images, the sphere position is measured using the `imfindcircles' function in
MATLAB. The image is binarized with a threshold of intensity between the
bright bulk and dark free surface intensity value. These binary values are
interpolated along a circular contour and filtered to keep large regions. The
transition between dark and light regions along this contour are used to
identify the position of the contact line between the sphere and free surface.

\section{\label{ExpResults} Experimental results}

First we will present a typical trial which demonstrates the main features of the drag profile as a function of height for the ascending and descending sphere. Next, we present experimental results that support a proposed mechanism for the observed increased drag. Then, we proceed to study the influence of wettability, flow speed, and sphere size.

An important reference for all of these cases is that of the free stream drag for single-phase (fully immersed) uniform flow.  The experiments presented here are in the subcritical range of Reynolds number preceding the onset of the drag crisis ($Re = 2-4 \times 10^{3}$) where $C_{D,0}$ is relatively constant ($C_{D,0}\approx0.4$) and also nearly independent of surface roughness. The nominal value of $C_{D,0}(Re)$ for fully immersed spheres used in this work is taken from established empirical results for this canonical problem \cite{morrison2013introduction}.

\begin{figure*}
\includegraphics[width=\textwidth]{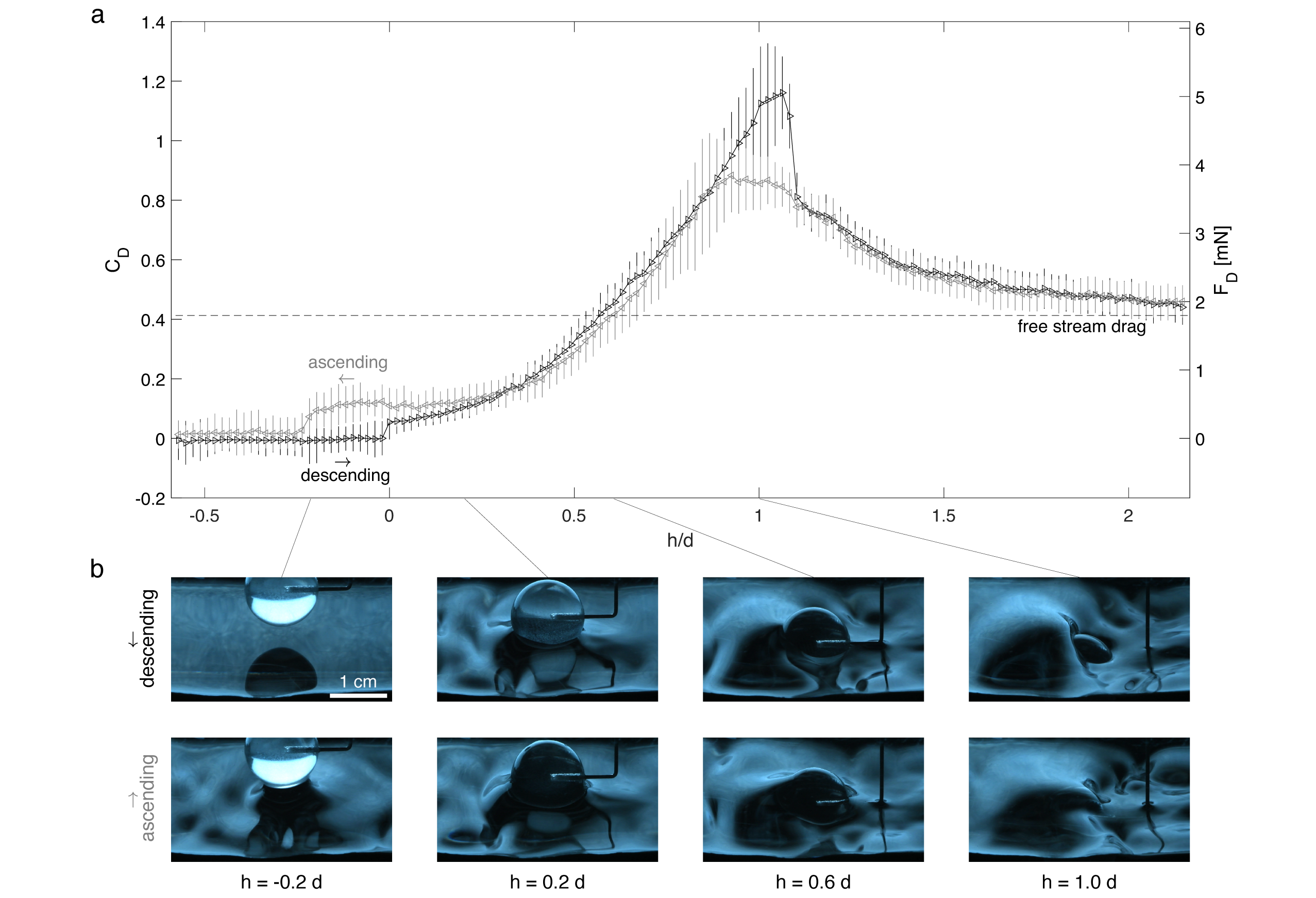}
\caption{\label{Single}(a) The measured sphere drag from a typical single trial. Acrylic sphere
  with $d = 12.7$ mm, $U = 26.2$ cm/s. Error bars represent
  the standard deviation of the unsteady force fluctuations. (b) Oblique photos at
  different submergence depths $h$ for the descending and ascending case.
  Submergence depths of $h = - 0.2 d$ and $h = 1.0 d$ represent regions of
  strong hysteresis.}
\end{figure*}

We begin with data obtained from a typical single trial (Figure \ref{Single}) where drag is measured on an acrylic
sphere with $d = 12.7$ mm, $U = 26.2$ cm/s as it both descends into and ascends back out of the flume.  The dimensional drag force on
the sphere is denoted $F_D$, and the coefficient of drag as $C_D$.  Error bars denote the standard deviation of measured unsteady force fluctuations.

The drag attains a maximum value when partially immersed that significantly
exceeds the free stream value. In the case considered here, this drag
reaches a value of nearly three times that of the equivalent fully immersed case.  Once the bottom of the sphere is submerged approximately two diameters below the undisturbed free surface height, the steady free stream drag value ($C_D \approx 0.4$) as expected from the literature is recovered.

Furthermore, there is an apparent hysteresis associated with the direction of
passage through the free surface.  The largest measured drag is associated
with the descending sphere \--- this occurs at the point just prior to the sphere becoming completely wetted.  On the return ascent, the peak drag is lesser, as the sphere is prewetted.  This result suggests that the solid wettability plays an
important role at these scales, and these effects are further interrogated in Section \ref{WettabilityEffects}.

There is an additional hysteresis in the region of small submergence depths, with a
sharp drag increase associated with the initial wetting of the sphere and spontaneous capillary rise. In the
ascending case, this small additional drag persists even as the sphere is retracted
above the equilibrium height of the free surface.

The hysteretic effects documented here are not due
to contact angle hysteresis, but rather to a bistability of immersed and
partially wetted states near points of topological transition of the interface (i.e. near $h/d=0, 1$).  A similar hysteresis of the equilibrium free surface profile can be observed even in the case of no flow for any finite equilibrium contact angle $0<\theta<\pi$. This distinction is perhaps most clearly demonstrated in
the left-most panels of Figure \ref{Single}(b) corresponding to $h = - 0.2 d$: for the
descending case, the sphere has not made contact with the water and does not
have a contact line.

\begin{figure*}
\includegraphics[width=\textwidth]{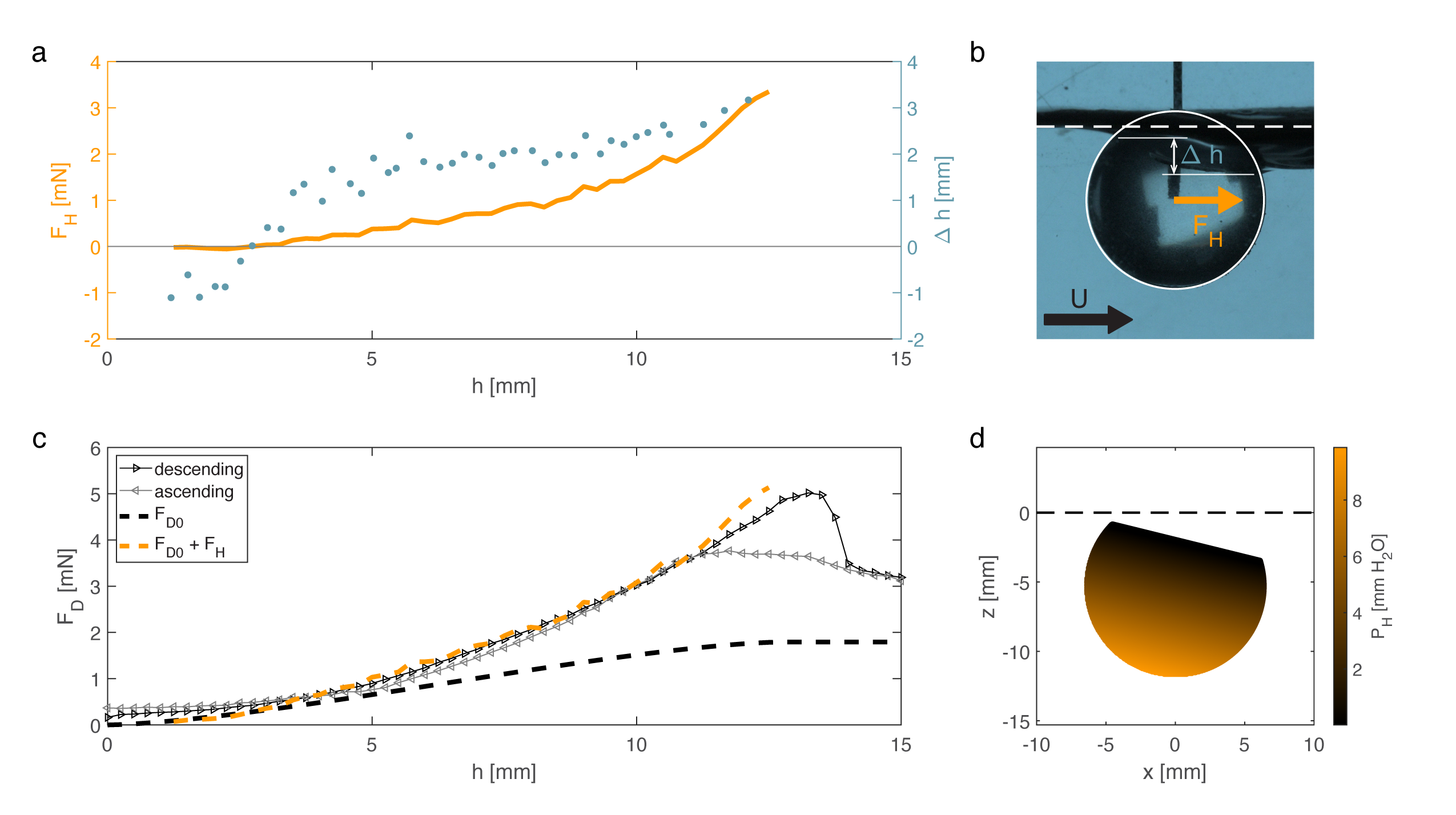}
\caption{\label{hydrostatic}(a) Plot of the estimated force due to hydrostatic effects $F_H$
  and the difference in height $\Delta h$ between the upstream and downstream
  contact points as measured experimentally. (b) Silhouette photo of the free
  surface used to measure $\Delta h$, with direction of estimated force $F_H$
  and flow $U$. (c) Plot of mean drag for the acrylic sphere ($U = 26.2$
  cm/s, $d = 12.7$ mm) in the descending and ascending
  directions, the estimated baseline form drag $F_{D, 0}$ derived from the free-stream drag
  scaled by the projected submerged area, estimated hydrostatic force $F_H$,
  and the sum of $F_{D, 0}$ and $F_H$. (d) The estimated hydrostatic pressure
  used to calculate the hydrostatic drag contribution $F_H$.}
\end{figure*}

\subsection{\label{HydrostaticEffects}Hydrostatic resistance}

Our data presented in Figure \ref{Single} demonstrates a substantial increase in drag due to the presence of the free surface.  In this section, we rationalize this additional drag by appealing to similar hydrostatic arguments as described by Teigen and Chaplin \cite{chaplin2003steady} for the case of the surface-piercing cylinder.

The Froude number characterizes the influence of inertial forces relative to
those of gravity. In particular, for a vertical surface of area $A$ and vertical length scale $L$,
the square of $Fr$ is proportional to the ratio
of forces from inertia ($\frac{1}{2} \rho U^2 A$) to those from hydrostatics ($\rho g L A$).

Our experiments are for $Fr \sim O (1)$, which suggests that
gravitational effects may contribute to our measurements.

Following Chaplin and Teigen \cite{chaplin2003steady}, we similarly postulate that the asymmetric hydrostatic pressure associated with the
disturbed free surface profile contributes significantly to the observed drag
increase.  To test this hypothesis, we measure the silhouette of the free
surface from the side using a Schlieren technique (see Section \ref{FreeSurfaceSilhouetteMeasurement}). We are able
to estimate the approximate hydrostatic contribution to the total drag by tracking the upstream and downstream position
of the free surface relative to the sphere.  The relative height difference of the upstream and downstream contact points is given as $\Delta h$. Away from the smallest submergence depths, $\Delta h$ takes an approximately constant value of $~2-3$ mm as shown in Figure \ref{hydrostatic}(a,b) and is consistent for both the ascending and descending cases when partially immersed.
We assume the contact line is
defined by the smallest circular path on the sphere connecting the measured
contact points furthest upstream and downstream. From this assumption, we
define the hydrostatic pressure on the sphere (Figure \ref{hydrostatic}(d)) by measuring the vertical distance to the plane containing this contact line, which is then integrated and projected in the horizontal direction to calculate
a hydrostatic drag contribution $F_H$. This pressure integral is similar to the approach of Havelock, where a prescribed pressure yields an expression for the surface height predicted using potential flow theory, and the resulting integrand is linearized and integrated over the entire surface. In contrast, we directly measure the free surface, derive the pressure field, and perform a full non-linear integral along the part of the free surface in contact with the sphere. 
To estimate the contribution of form drag on the sphere in the
absence of free surface effects, we scale the empirical free stream drag by
the fractional immersed projected area, defined as:
\begin{eqnarray*}
  F_{D, 0} & = & \frac{1}{2} C_{D,0} \rho U^2 \pi \left({\frac{d}{2}}\right)^2 \chi(h/d)
\end{eqnarray*}
where $\chi$ is the fraction of the projected area submerged relative to the
quiescent water level and ranges from 0 (fully removed) to 1 (fully immersed).  We additionally assume that $C_{D,0}$ is independent of $\chi$ and is equal to the corresponding single-phase value.

The additional estimated drag due to hydrostatics, $F_H$, is on the same order of
magnitude as the fractionally scaled form drag $F_{D, 0}$. These two forces, along with
their sum, are overlayed in Figure \ref{hydrostatic} with the mean drag measured
experimentally for the equivalent conditions (acrylic sphere, $d = 12.7$
mm, $U = 26.2$ cm/s) with very good agreement.  Although the excellent agreement may be fortuitous in this case given the number of assumptions required in the prediction, it is nevertheless evident that the hydrostatic contribution is a viable dominant mechanism for the observed drag increase in our partially immersed experiments.  Although more work across operating conditions is
needed to validate the simple model proposed herein, we expect that
hydrostatic forces due to asymmetric free-surface-solid interactions are
relevant in more general scenarios whenever $Fr \sim O (1)$.  As later demonstrated in Section \ref{SpeedVariations}, the measured drag increase does not change significantly whether one is above or below the critical wave speed, suggesting that wave drag due to radiated wave energy is not the dominant free-surface drag mechanism in our experiments.

\subsection{\label{WettabilityEffects}Influence of solid wettability}

\begin{figure*}
\includegraphics[width=\textwidth]{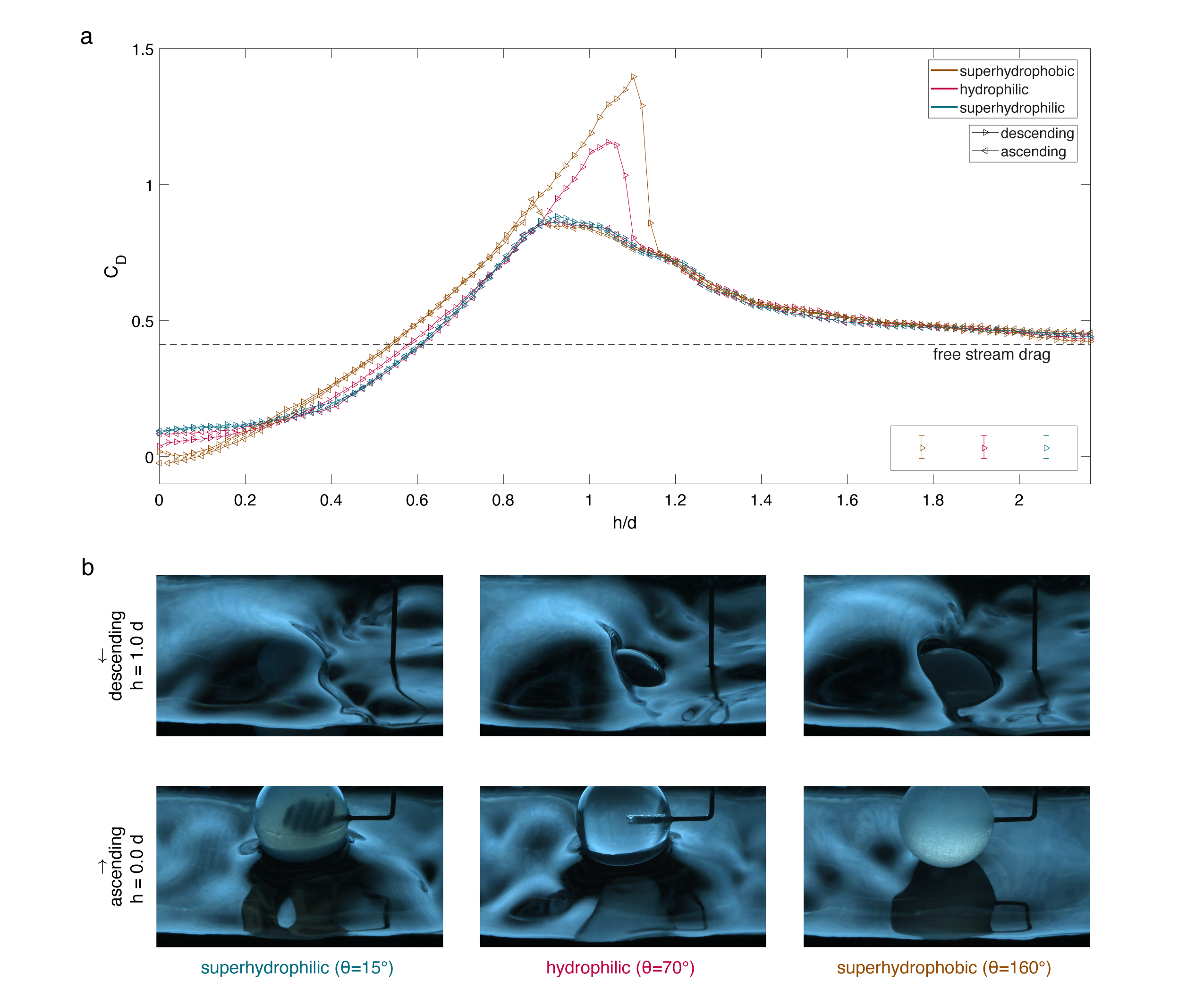}
\caption{\label{Coatings}(a) Comparison of mean drag for three otherwise identical conditions ($d = 12.7$ mm, $U = 26.2$ cm/s), varying only the equilibrium contact angle ($\theta=15^{\circ},70^{\circ},160^{\circ}$).
  Inset in lower right shows estimated uncertainty for each case. (b) Photos for the three spheres at
  $h = 1.0 d$ for the descending case and $h = 0.0 d$ for the ascending case.}
\end{figure*}
We tested the effect of solid wettability on the measured drag at all depths by performing experiments on otherwise identical spheres with three
different equilibrium contact angles: agar ($15^{\circ}$), acrylic ($70^{\circ}$) and NeverWet coated
acrylic ($160^{\circ}$). The results from the three cases are overlaid in Figure \ref{Coatings}.  The drag curves demonstrate the significant influence of
wettability on the drag characteristics near the peak: for the superhydrophilic (agar) case, the large force peak associated with the incomplete wetting in the acrylic case is absent, and the
hysteretic effect in this region is subdued.  On the contrary, the superhydrophobic
(NeverWet) case exhibits a much larger force maximum, corresponding to the
delayed transition to full wetting which now occurs at a larger submergence depth. It is important to note that, as discussed in the previous section, this significant hysteretic effect is not due to contact line hysteresis, which is very small for this coating \cite{weisensee2017droplet}.  Typically, superhydrophobic coatings are associated with drag {\it reduction}, however in this case, the high contact angle inhibits complete wetting and increases the drag substantially near full submergence.  For superhydrophilic spheres
at submergence depths near $h = 1.0 d$, their promotion of wetting causes a
smaller overall disturbance of the free surface and a smaller resulting drag.

For measurements in the region of small submersion depths,
the superhydrophobic (NeverWet) sphere does not show a sharp drag increase
associated with spontaneous wetting as in the acrylic and agar cases, and the increased drag
associated with wetting when the sphere is retracted above the equilibrium
water height is similarly absent. For superhydrophobic spheres at small submergence
depths, their reluctance to wet causes a smaller overall disturbance of the free
surface and a smaller resulting drag.

In summary, to reduce drag when the sphere is near full submergence, it is beneficial to promote wetting. To reduce drag for small submersion depths, one should inhibit wetting. Ultimately for a sphere at the interface to have minimal drag and hysteresis
over all submergence depths, these results suggest a hybrid solution: a superhydrophobic
bottom and superhydrophilic top.

\subsection{\label{SpeedVariations}Influence of flow speed}
\begin{figure*}
\includegraphics[width=\textwidth]{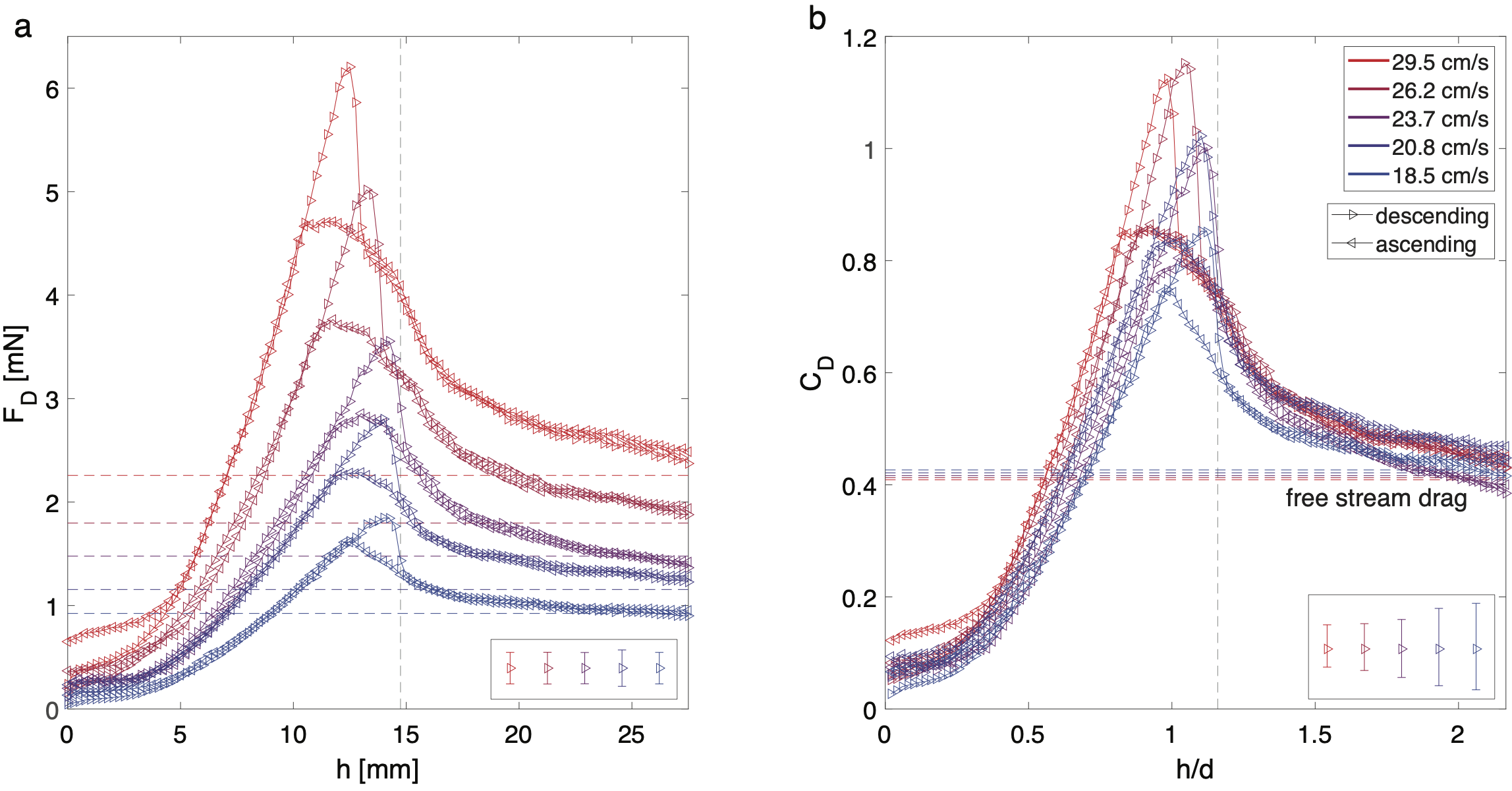}
\caption{\label{Speed}Mean drag for the acrylic sphere ($d = 12.7$ mm) at five
  speeds ($U = 18.5, 20.8, 23.7, 26.2, 29.5$ cm/s) shown (a) dimensionally and (b)
  nondimensionally. Inset in lower right shows
  mean estimated uncertainty.  The vertical dashed lines indicate the measured depth at which an equivalent descending sphere in the absence of flow undergoes complete immersion.}
\end{figure*}
Next we move to explore the influence of flow speed on the drag force within the range of our flume's capability.  For an acrylic sphere with $d = 12.7$ mm, we measured the drag at 5
different speeds $U$ ranging from 18.5-29.5 cm/s, corresponding to $Re
= 2.3 - 3.7 \times 10^3$ ($C_{D,0}\approx 0.4$) and $Fr = .52 - .84$.   The results, plotted in Figure \ref{Speed}, nearly collapse under inertial rescaling.
The main features initially documented in Figure \ref{Single} are robust across the
different speeds tested, including a force maximum that significantly exceeds the
fully immersed free stream drag and significant hysteretic effects near the peak.  One trend, however, is the evolution of the critical submergence depth for
which full wetting occurs in the descending case, again associated with a sharp
decrease in the measured drag.  In particular, the faster flows (higher $Fr$) allow for complete wetting to occur at a smaller submergence depth.  For the slower flow cases, this transition occurs for $h
> d$ and tends to the corresponding wetting transition in the case of no flow as indicated by the vertical dashed lines in Figure \ref{Speed}.
The exact value it tends to is expected to be strongly associated with the solid wettability. 

Furthermore, we observe a similar increase in drag coefficient over the range of speeds considered here which spans the critical minimum wavespeed for capillary-gravity waves (23 cm/s), suggesting the
onset of wave radiation as might be predicted by pure potential flow arguments is not a significant contributor to the excess drag measured in the present experiments.

\subsection{\label{SizeVariations}Influence of sphere diameter}
\begin{figure*}
\includegraphics[width=\textwidth]{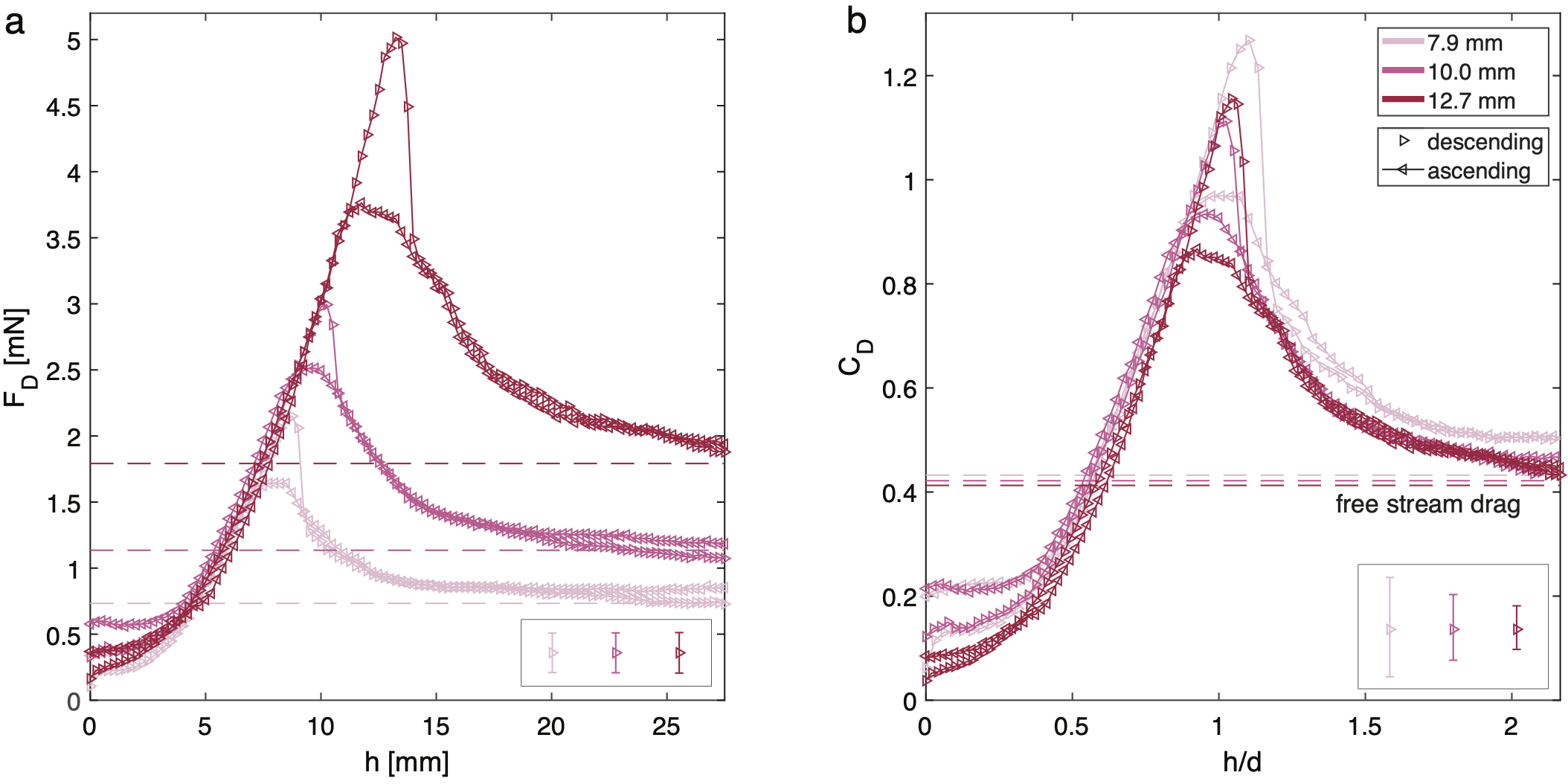}
\caption{\label{Size}Mean drag for acrylic spheres of three different sizes ($U = 26.2 \
  \mathrm{cm/s}, d = 7.9, 10.0, 12.7 \ \mathrm{mm}$)  (a) dimensionally and (b)
  nondimensionally. Inset in lower right shows mean estimated
  uncertainty.}
\end{figure*}

Lastly, we investigate the role of sphere diameter on the observed drag characteristics. Again, the range tested herein is only modest due to the constraints of the experimental platform.  Significantly smaller spheres would require a more accurate force sensor, whereas significantly larger spheres would require a larger test section to avoid blockage effects.  For a fixed flow speed $U = 26.2$ cm/s, we studied the drag on
acrylic spheres of diameters 7.9 mm, 10 mm, and 12.7 mm, corresponding to
$Re = 2.1 - 3.3 \times 10^3$ ($C_{D,0}\approx 0.4$) and $Fr = .74 - .94$. The drag
curves again collapse reasonably well when non-dimensionalized under inertial scaling. Once again, as initially observed in Figure \ref{Single}, the maximum force is significantly larger than the free
stream drag, and hysteretic effects cause significant differences in the
measured drag at small submergence depths and near the force peak. Some deviations do occur in the hysteretic regions between cases, but the uncertainty from the force sensor in this study is relatively large in comparison to the observed differences in drag.

\section{\label{NumericalSimulations} Numerical simulations}
\begin{figure*}
\includegraphics[width=\textwidth]{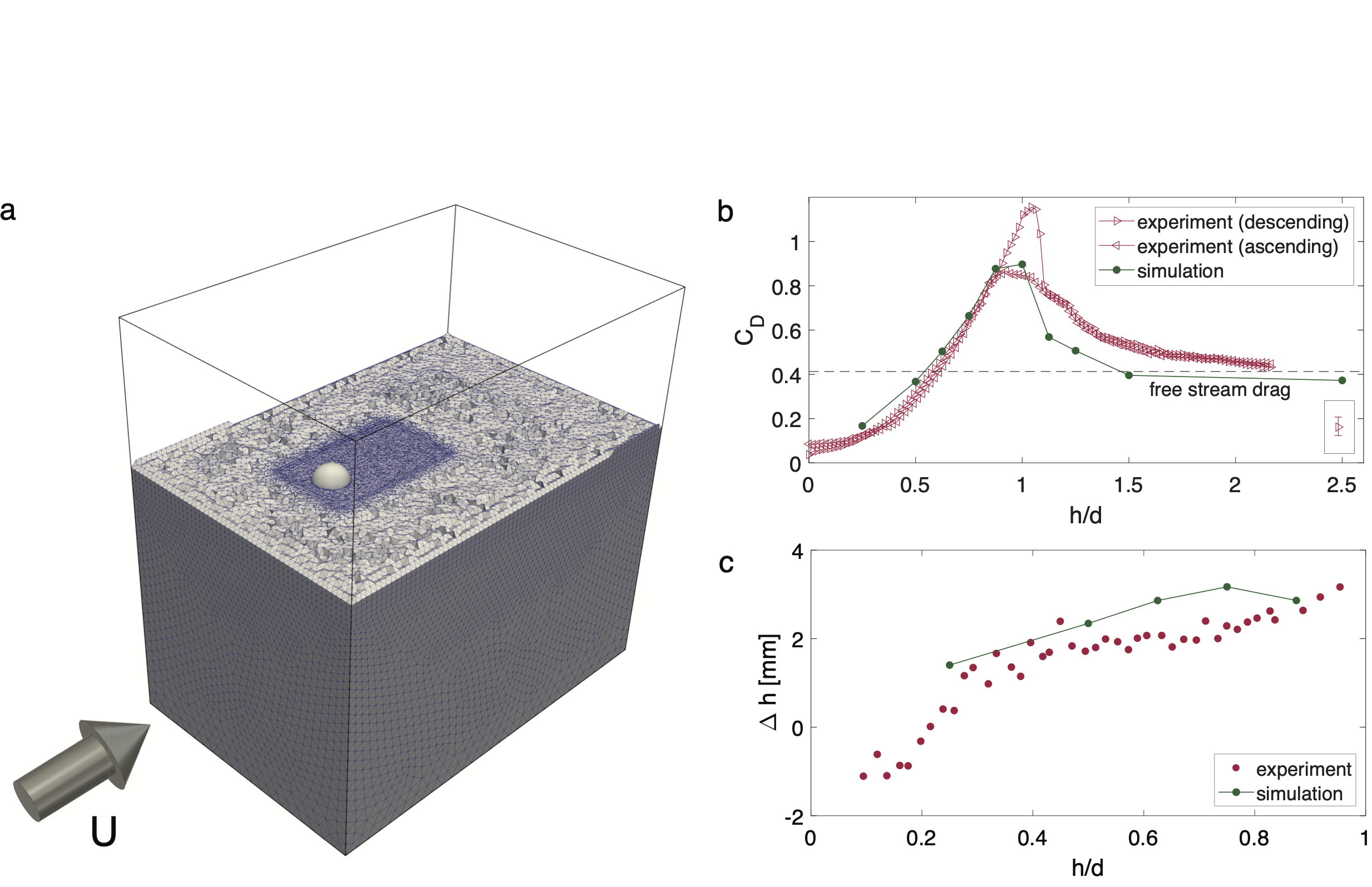}
\caption{\label{fig:mesh} (a) Element distribution in the mesh where the submergence depth is $h=0.625d$. The outer box uses elements of size 5.0 mm where as the elements around the sphere and free surface are refined for better resolution to 0.5 mm and 2.0 mm, respectively. Numerical results for (b) drag and (c) upstream/downstream height difference in a $10 \times 10$ cm cross section compared with experimental results for an acrylic sphere ($U = 26.2 \
  \mathrm{cm/s}, d = 12.7 \ \mathrm{mm}, \gamma = 72.2 \  \mathrm{mN/m} $) }
\end{figure*}

To complement the experiments, we perform a series of numerical simulations of free surface flow past a sphere at various immersion depths.  The numerical study focuses on the influence of surface tension which is difficult to control as an independent parameter in the experiment.

\subsection{\label{NumericalMethods} Numerical methods}

The intermediate energies and length scales considered in this work will lead to the co-existence of laminar, transitional, and turbulent flow regions in the domain of interest. For this reason, we adopt a variational multiscale (VMS) formulation \cite{bazilevs2007variational} of fluid dynamics, which was shown to possess excellent accuracy and stability characteristics across a wide range of Reynolds numbers. For higher Reynolds number flows, VMS was shown to behave similarly to Large-Eddy Simulation (LES) models of turbulence, but without the reliance on empirical eddy viscosities that do not perform well in transitional flows. The VMS makes use of a weak form of the Navier–Stokes equations and may be discretized with Finite Element~\cite{hsu2012wind,hsu2014finite,bazilevs2014computation,bazilevs2015ale,xu2016tetrahedral,yan2017new,zhao2022enriched} or Isogeometric Analysis methods~\cite{hsu2011high,van2017isogeometric,bazilevs2013isogeometric}. To avoid costly boundary-layer resolution typically required for LES computations, the VMS formulation is augmented with weakly-enforced no-slip boundary conditions~\cite{bazilevs2007weak, Bazilevs07c}, which may be thought of as a near-wall model.

The level-set approach~\cite{sussman1994level,osher1988fronts} is adopted to track the evolution of the free surface. The aerodynamics and hydrodynamics are governed by the Navier-Stokes equations of incompressible two-fluid flow, in which the fluid density and viscosity are evaluated with the aid of a level-set function whose zero level is passively convected with the flow and used to define the air-water interface. In addition, we utilize a geometry-based re-distancing technique to maintain the signed-distance property of the level-set field~\cite{zhu2021mixed}. The free-surface framework is augmented to include surface tension using a balanced-force surface tension model~\cite{brackbill1992continuum, zhao2021variational}. The combined free-surface flow framework was extensively validated in~\cite{akkerman2011isogeometric, yan2016computational,yan2017free,yan2018fully, yan2019isogeometric, zhu2020immersogeometric} for similar problems considered in this paper and is employed here to carry out the computations that accompany the experiments presented in the prior sections.

The simulations use $U= 26.2 \ \mathrm{cm/s}$ and $d=12.7\ \mathrm{mm}$. Ten immersion depths are investigated ($0.25d$, $0.5d$, $0.625d$, $0.75d$, $0.875d$, $d$, $1.125d$, $1.25d$, $1.5d$, and $2.5d$). The simulated domain is a box of $15 \ \mathrm{cm}\times 10 \ \mathrm{cm}\times 15 \ \mathrm{cm}$, and the sphere is placed $4 \ \mathrm{cm}$ downstream from the inlet. The depth of the water phase is $10 \ \mathrm{cm}$, leading to a $10 \ \mathrm{cm}\times10 \ \mathrm{cm}$ cross-section.

The simulations utilize unstructured linear tetrahedral elements. Two additional refinement boxes are generated to capture the wake behind the sphere and free surface deformation. Element sizes are 0.5 mm around the sphere, 2.0 mm near the free surface, and 5.0 mm in the outer box.
Figure~\ref{fig:mesh} (a) presents a snapshot of the mesh in the center plane with a submergence depth of $h=0.625d$. In the simulations, we apply a uniform inflow velocity $U= 26.2$ cm/s at the inlet and no penetration boundary condition on the lateral walls. Hydrostatic equilibrium pressure is applied at the outlet. In addition to the physical surface tension coefficient of $\gamma= 72.2$ mN/m, we also conduct simulations with $\gamma= 36.1$ mN/m, and 0 mN/m. The time step size is set to $5\times10^{-4} \ \mathrm{s}$ for all of the simulations. We simulate all of the cases until the results reach a statistically stationary stage. The data in the fully developed stage is used to obtain the mean drag coefficients from the simulation.

\begin{figure*}
\includegraphics[width=\textwidth]{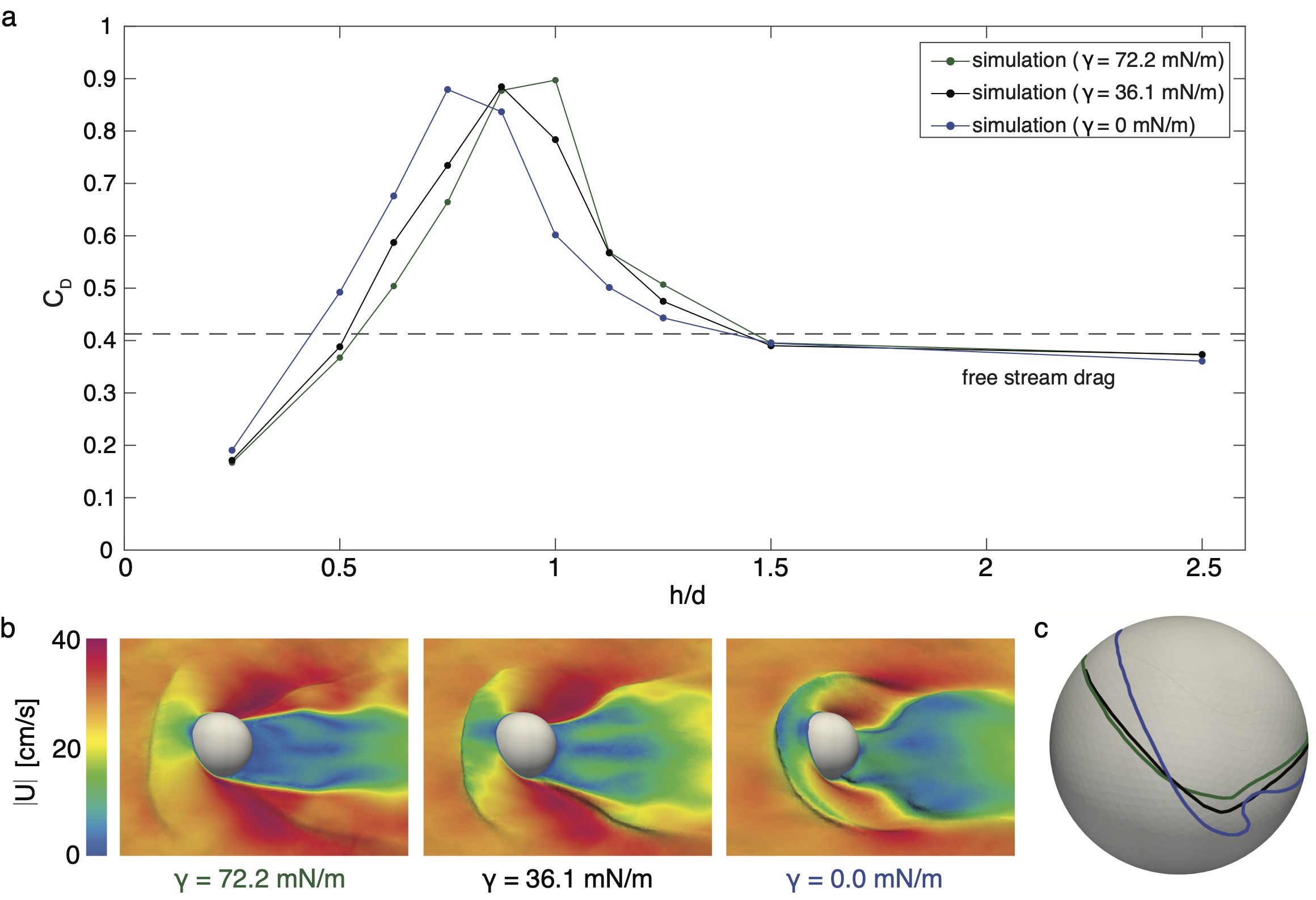}
\caption{\label{NumComparisonST} (a) Numerical results for drag on a sphere of diameter $d = 12.7 \  \mathrm{mm}$ in  uniform inlet flow with speed $U = 26.2$ cm/s near a free surface with varying surface tension coefficient ($\gamma = 72.2, \ 36.1, \ 0.0 \ \mathrm{mN/m}$). Time-averaged (b) free surface and surface flow speed speed and (c) contact line position under the various surface tensions in a partially immersed case corresponding to submergence depth $h=0.625d$.}
\end{figure*}

\subsection{\label{NumericalResults} Results}
\textcolor{black}{To assess the influence of any possible blockage effects in our experiment, we first run simulations at two submergence depths ($h=0.5d$ and $1.0d$) using a smaller $5 \ \mathrm{cm} \times 5 \ \mathrm{cm}$ inlet flow cross-section as in the experiment. These results are compared with those of the $10 \ \mathrm{cm} \times 10 \ \mathrm{cm}$ cross-section simulations in Table~\ref{tab:cross-section}. The limited discrepancy suggests any blockage effects are negligible.}
\begin{table}
    \centering
    \begin{tabular}{|c|c|c|}
    \hline
    Depth ($h$) & $10 \ \mathrm{cm}\times10 \ \mathrm{cm}$ & $5 \ \mathrm{cm}\times 5 \ \mathrm{cm}$ \\
    \hline
    $0.5d$ & 0.3673 & 0.3749\\
    \hline
    $d$ & 0.8971 &0.8847 \\
    \hline
    \end{tabular}
    \caption{Time-averaged drag coefficients comparison between $5 \ \mathrm{cm}\times 5 \ \mathrm{cm}$ and $10 \ \mathrm{cm}\times 10 \ \mathrm{cm}$ flow inlet cross sections at two submergence depths for sphere diameter $d = 12.7 \  \mathrm{mm}$, flow speed $U = 26.2$ cm/s, and surface tension $\gamma = 72.2\ \mathrm{mN/m}$).  The depth $h=0.5d$ corresponds to a partially immersed case whereas the depth $h=d$ corresponds to a fully immersed case.}
    \label{tab:cross-section}
\end{table}
\textcolor{black}{Figure~\ref{fig:mesh}(b) compares the experimental and numerical prediction of drag coefficients using the physical surface tension coefficient $72.2$ mN/m.  The results show excellent agreement for the partially immersed cases and similarly predict a force peak that can significantly exceed the nominal free stream value.  Near the peak, the numerical prediction of the drag peak is closer to the ascending cases which were shown in experiment to be significantly less influenced by the solid wettability. The peak happens approximately when the water moves fully across spheres from above.  In the simulation, once the water reaches the downstream side of the sphere, it will immediately merge into the bulk flow. Due to the lack of contact line models in the present simulation, the water does not have enough support to resist gravity, which is similar to the phenomenon observed in experiments for the superhydrophilic sphere that also minimally impedes contact line progression.  In that case, the descending and ascending cases are nearly indistinguishable near the peak.  Therefore, the simulated prediction of the drag peak is closer to that of the ascending cases.  For the fully submerged cases, the measured drag is greater than the numerical one likely due to the presence of weak free stream turbulence in experiments. The drag of a fully submerged sphere mainly results from low-pressure regions related to flow separation. The remaining turbulence in the experimental inflow triggers flow separation and increases the low-pressure area behind the spheres. The different contributions to the drag between the partially and fully immersed cases will be further studied in what follows.}

Figure~\ref{fig:mesh}(c) shows the difference in height between the upstream and downstream contact points, comparing the experimental measurements (also shown in Figure 3(a)) and the computational results. The height difference $\Delta h$ is plotted as a function of the submergence depth $h$. To compute $\Delta h$, the time-averaged free surface profile is projected onto the tank side surface in the span-wise direction and the resulting curve is intersected with the circle of diameter $d$ that defines the silhouette of the sphere. The intersection points upstream and downstream of the sphere define the positions of the contact points and the difference between the two positions defines $\Delta h$. As can be observed from the figure, the experimental and computational results match very well for lower submergence depths. However, higher submergence depths give more discrepancy in the results, which is also likely attributable to the lack of explicit modeling of the contact line in the free-surface flow simulations. While the current modeling approach can accurately predict the drag force increase relative to the free-stream case, predicting the finer-scale features of the free surface near the solid object will require the corresponding enhancements in the modeling framework, which we plan to undertake in the future.

 Figure~\ref{NumComparisonST}(a) presents the drag coefficient versus submersion depth with three surface tension coefficients. The drag coefficient of partially submerged spheres is negatively correlated with surface tension while positively correlated for the fully submerged case.  When the sphere is exposed to the air, the air phase has a minimal contribution to the drag compared to the water due to its low density.  Surface tension affects the additional drag by modulating the shape of the free surface and the water height difference between the front and back of the sphere. The strength of surface tension represents the capability of generally reducing deformations of the air-water interface. Figure~\ref{NumComparisonST}(b) depicts the mean free surface deformation under different surface tensions for a partially immersed case ($h=0.625d$). Apart from small wrinkles, the presence of surface tension suppresses water climbing at the upstream side and dropping at the downstream side, thus contributing to a lower drag as might be anticipated by the hydrostatic mechanism described in Section ~\ref{HydrostaticEffects}.  Figure~\ref{NumComparisonST}(c), which shows where the time-averaged interface contacts the sphere, further validates this point. 
 
\begin{figure*}
\includegraphics[width=\textwidth]{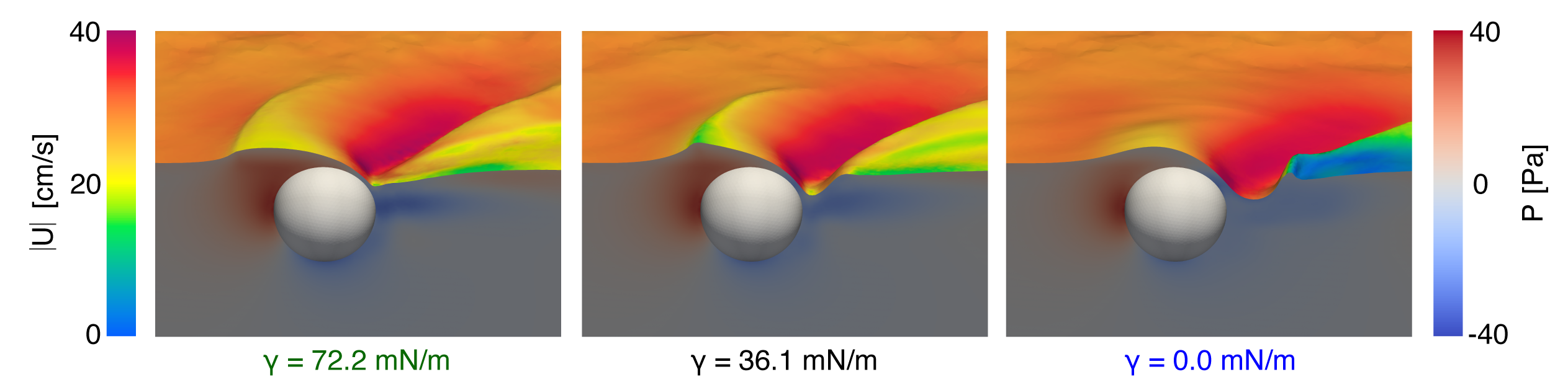}
\caption{\label{fig:deformation-full} The time-averaged free surface and pressure distribution in the water phase along the sphere's central plane under various surface tension with submergence depth $h=d$. From left to right, $\gamma=0$, $36.1$, and $72.2 \ \mathrm{mN/m}$, respectively.}
\end{figure*}

Figure~\ref{fig:deformation-full} shows how surface tension influences the free surface shape and pressure distribution in a case where the spheres are fully immersed ($h=1.0d$). As mentioned previously, we attribute the drag of a fully immersed sphere primarily to form drag which is related to flow separation.  The flattening of the surface behind the sphere due to surface tension in this case leads to an earlier separation point, and thus a larger suction region on the downstream side.  Increased surface tension thus leads to an increased drag in this regime, in contrast to the partially immersed regime where the trend is opposite.

\section{\label{Discussion} Discussion}

Although conceptually similar to drag on a sphere in single-phase flow, the
significant surface deformations introduced by an obstacle penetrating a free surface preclude any attempt at direct extrapolation from single-phase drag measurements.  The subtle interplay between the presence of an obstacle, its concomitant disturbance in the fluid flow, and the resulting asymmetric free surface play an essential role in determining the ultimate drag experienced by the body.

To the knowledge of the authors, there is only one prior work that
has studied the drag on a fixed, surface-piercing sphere as a function of depth.
James \cite{james2016search} explored the drag on towed, fixed spheres of diameters $d = 12.5$
and $22.5$ cm, for four submergence depths of $h = .25d, .40d, .50d,$ and $.75d.$, 
for $U = 60 - 370$ cm/s.  Due to the presence of the free surface, they observe a non-monotonic drag enhancement, in one case reaching a factor of four times the free stream drag. They observe a drag maximum as a function of $Fr$ and a corresponding decrease in drag occurring when the upstream rise
height reaches the top of the sphere. In particular, they propose a similar
model as Chaplin and Teigan \cite{chaplin2003steady}, where the rise height scales as $Fr^2$, and
successfully fit experimental observations of this observed Froude transition with
this scaling after applying a correction factor.

Our experiments exhibit a similar Froude transition corresponding to the observed
force maximum just prior to full immersion.  However, the transition as documented here is strongly hysteretic and can occur for submergence depths larger than $1.0 d$ due to the significance of solid wettability in our case.  Furthermore, we have demonstrated that an appreciable drag enhancement is present regardless of whether one operates above or below the minimum capillary-gravity wave speed.

The enhanced drag associated with the presence of the free surface and its sensitive dependence on immersion depth and wettability will be useful in informing future designs of interfacial ``micro''-robots \cite{yuan2012bio,chen2018controllable,timm2021remotely,rhee2022surferbot} and increasing our understanding of propulsion and drag of surface-dwelling insects \cite{bush2006walking}.  It may also prove insightful to other applications where particles obliquely impact a free surface at shallow angles, as in the case of particulate capture by raindrops \cite{speirs2023capture} or in natural settings where surface-piercing objects appear, such as sediment transport in shallow streams \cite{lamb2017direct} and other free-surface particulate flows \cite{singh2006free,zhou2022drag}.

In this work, we have shown that the additional drag can be accounted for by considering the asymmetric hydrostatic force contribution resulting from the difference in fluid height between the upstream and downstream side of the sphere.  This run-up effect was studied by Hay \cite{hay1947} and later Chaplin and Teigan \cite{chaplin2003steady} in the context of
submerged cylinders. The height of the run-up relative to the quiescent
water level, $h_{rise}$, can be estimated from Bernoulli's equation as
\begin{eqnarray*}
  h_{rise} / d & = & \frac{1}{2} Fr^2 .
\end{eqnarray*}
They directly measured the run-up height as a function of $Re
/ Fr$, which represents the ratio of gravitational forces to viscous
forces, and demonstrate that the $Fr^2$ scaling is recovered in the limit of high $Re / Fr$.  To relate this to the excess force on the cylinder, Chaplin and Teigan proposed a model similar to ours which estimates the hydrostatic pressure on the
upstream face of the cylinder due to run-up.  The additional hydrostatic contribution is only considered {\it above} the quiescent water
line however. They use this pressure estimate to predict an additional contribution to the drag coefficient that should also scale like $Fr^2$ for low $Fr$. All of their experiments were performed above the minimum capillary-gravity wave speed and at large scales where capillarity and solid wettability are less likely to play a significant role.

A number of previous related studies have presented experiments and theory focused specifically on the onset of wave drag \cite{raphael1996,browaeys2001capillary,burghelea2001,burghelea2002,pham2005critical,merrer2011,benzaquen2011wave} above the minimum capillary-gravity wave speed as predicted by potential flow arguments. One might thus be tempted to believe that free surface effects are unlikely to play a significant role in determining the drag below such speeds. While Richard et al. \cite{richard1999capillary} predict some drag contribution below the threshold when weak viscous effects are introduced, a substantial drag increase is still anticipated when crossing the threshold. While our results do not intend to rule out such a contribution to the total drag, they do demonstrate that, in practice, interfacial drag mechanisms are highly significant independent of whether one operates below or above this theoretical threshold. 

To conclude, we have investigated the drag on a partially immersed centimetric sphere moving along an air-water interface.  Our direct measurements and computational predictions demonstrate a significant drag increase over the equivalent single-phase flow case.  A simple model is proposed to estimate the extra drag from hydrostatics using the measured difference in fluid height between the upstream and downstream as input.  While flow speed and sphere size have a relatively minor influence on the drag coefficients for the regime studied herein, solid wettability is shown to play a relatively pronounced role at such small scales.  While this work has primarily focused on the influence of submergence depth and surface wettability, in future work we plan to significantly broaden the parameter space considered here in order to more systematically test the proposed model and better characterize the sphere drag and Froude crisis over a wider range of Reynolds, Froude, and Bond numbers.  Additionally, vertical forces (lift) should be considered in future work, which are highly relevant to bodies that freely float or self-propel at the interface.  Lastly, the principal phenomena revealed here will be explored for classes of more general body geometries and arrays of multiple bodies.
\newline

Data presented in this work is publicly available at the following url:
\url{https://github.com/harrislab-brown/FreeSurfaceSphereDrag}.

\begin{acknowledgments}
This research was supported through the ONR Grant No. N00014-21-1-2670. The authors wish to thank Drs. Jesse Belden, Nathan Speirs, George Loubimov, and Kristina Kamensky of the Naval Undersea Warfare Center (NUWC), Newport for providing valuable suggestions and feedback for this research. R.H., E.S., and D.M.H. additionally thank Prof. Kenny Breuer for use of the laser to perform the flume characterization PIV measurements.
\end{acknowledgments}

\bibliography{apssamp}

\providecommand{\noopsort}[1]{}\providecommand{\singleletter}[1]{#1}%
\begin{thebibliography}{69}%
\makeatletter
\providecommand \@ifxundefined [1]{%
 \@ifx{#1\undefined}
}%
\providecommand \@ifnum [1]{%
 \ifnum #1\expandafter \@firstoftwo
 \else \expandafter \@secondoftwo
 \fi
}%
\providecommand \@ifx [1]{%
 \ifx #1\expandafter \@firstoftwo
 \else \expandafter \@secondoftwo
 \fi
}%
\providecommand \natexlab [1]{#1}%
\providecommand \enquote  [1]{``#1''}%
\providecommand \bibnamefont  [1]{#1}%
\providecommand \bibfnamefont [1]{#1}%
\providecommand \citenamefont [1]{#1}%
\providecommand \href@noop [0]{\@secondoftwo}%
\providecommand \href [0]{\begingroup \@sanitize@url \@href}%
\providecommand \@href[1]{\@@startlink{#1}\@@href}%
\providecommand \@@href[1]{\endgroup#1\@@endlink}%
\providecommand \@sanitize@url [0]{\catcode `\\12\catcode `\$12\catcode
  `\&12\catcode `\#12\catcode `\^12\catcode `\_12\catcode `\%12\relax}%
\providecommand \@@startlink[1]{}%
\providecommand \@@endlink[0]{}%
\providecommand \url  [0]{\begingroup\@sanitize@url \@url }%
\providecommand \@url [1]{\endgroup\@href {#1}{\urlprefix }}%
\providecommand \urlprefix  [0]{URL }%
\providecommand \Eprint [0]{\href }%
\providecommand \doibase [0]{https://doi.org/}%
\providecommand \selectlanguage [0]{\@gobble}%
\providecommand \bibinfo  [0]{\@secondoftwo}%
\providecommand \bibfield  [0]{\@secondoftwo}%
\providecommand \translation [1]{[#1]}%
\providecommand \BibitemOpen [0]{}%
\providecommand \bibitemStop [0]{}%
\providecommand \bibitemNoStop [0]{.\EOS\space}%
\providecommand \EOS [0]{\spacefactor3000\relax}%
\providecommand \BibitemShut  [1]{\csname bibitem#1\endcsname}%
\let\auto@bib@innerbib\@empty
\bibitem [{\citenamefont {Russell}(1839)}]{russell1839iii}%
  \BibitemOpen
  \bibfield  {author} {\bibinfo {author} {\bibfnamefont {J.~S.}\ \bibnamefont
  {Russell}},\ }\bibfield  {title} {\bibinfo {title} {Experimental researches
  into the laws of certain hydrodynamical phenomena that accompany the motion
  of floating bodies...},\ }\href@noop {} {\bibfield  {journal} {\bibinfo
  {journal} {Earth and Environmental Science Transactions of The Royal Society
  of Edinburgh}\ }\textbf {\bibinfo {volume} {14}} (\bibinfo {year}
  {1839})}\BibitemShut {NoStop}%
\bibitem [{\citenamefont {Reech}(1852)}]{reech1852course}%
  \BibitemOpen
  \bibfield  {author} {\bibinfo {author} {\bibfnamefont {F.}~\bibnamefont
  {Reech}},\ }\href@noop {} {\emph {\bibinfo {title} {Cours de m{\'e}canique
  d'apr{\`e}s la nature g{\'e}n{\'e}ralement flexible et {\'e}lastique des
  corps...}}}\ (\bibinfo  {publisher} {Carilian-Goeury et Vor Dalmont},\
  \bibinfo {year} {1852})\BibitemShut {NoStop}%
\bibitem [{\citenamefont {Froude}(1887)}]{froude1887}%
  \BibitemOpen
  \bibfield  {author} {\bibinfo {author} {\bibfnamefont {W.}~\bibnamefont
  {Froude}},\ }\bibfield  {title} {\bibinfo {title} {On experiments upon the
  effect produced on the wave making resistance of ships by length of parallel
  middle body},\ }\href@noop {} {\bibfield  {journal} {\bibinfo  {journal}
  {Transactions of the Institute of Naval Architects}\ }\textbf {\bibinfo
  {volume} {18}} (\bibinfo {year} {1887})}\BibitemShut {NoStop}%
\bibitem [{\citenamefont {Thomson}(1887)}]{thomson1887}%
  \BibitemOpen
  \bibfield  {author} {\bibinfo {author} {\bibfnamefont {W.}~\bibnamefont
  {Thomson}},\ }\bibfield  {title} {\bibinfo {title} {On ship waves},\
  }\href@noop {} {\bibfield  {journal} {\bibinfo  {journal} {Proceedings of the
  Institution of Mechanical Engineers}\ }\textbf {\bibinfo {volume} {38}}
  (\bibinfo {year} {1887})}\BibitemShut {NoStop}%
\bibitem [{\citenamefont {Lamb}(1913)}]{lamb1913}%
  \BibitemOpen
  \bibfield  {author} {\bibinfo {author} {\bibfnamefont {H.}~\bibnamefont
  {Lamb}},\ }\bibfield  {title} {\bibinfo {title} {On some cases of wave-motion
  on deep water},\ }\href@noop {} {\bibfield  {journal} {\bibinfo  {journal}
  {Annali di Matematica Pura ed Applicata}\ }\textbf {\bibinfo {volume} {21}}
  (\bibinfo {year} {1913})}\BibitemShut {NoStop}%
\bibitem [{\citenamefont {Havelock}(1917)}]{havelock1917some}%
  \BibitemOpen
  \bibfield  {author} {\bibinfo {author} {\bibfnamefont {T.~H.}\ \bibnamefont
  {Havelock}},\ }\bibfield  {title} {\bibinfo {title} {Some cases of wave
  motion due to a submerged obstacle},\ }\href@noop {} {\bibfield  {journal}
  {\bibinfo  {journal} {Proceedings of the Royal Society of London, Series A}\
  }\textbf {\bibinfo {volume} {93}} (\bibinfo {year} {1917})}\BibitemShut
  {NoStop}%
\bibitem [{\citenamefont {Newman}(1977)}]{newman1977marine}%
  \BibitemOpen
  \bibfield  {author} {\bibinfo {author} {\bibfnamefont {J.~N.}\ \bibnamefont
  {Newman}},\ }\href@noop {} {\emph {\bibinfo {title} {Marine hydrodynamics}}}\
  (\bibinfo  {publisher} {The MIT press},\ \bibinfo {year} {1977})\BibitemShut
  {NoStop}%
\bibitem [{\citenamefont {Dawson}(2014)}]{dawson2014investigation}%
  \BibitemOpen
  \bibfield  {author} {\bibinfo {author} {\bibfnamefont {E.}~\bibnamefont
  {Dawson}},\ }\emph {\bibinfo {title} {An investigation into the effects of
  submergence depth, speed and hull length-to-diameter ratio on the near
  surface operation of conventional submarines}},\ \href@noop {} {Ph.D.
  thesis},\ \bibinfo  {school} {University of Tasmania} (\bibinfo {year}
  {2014})\BibitemShut {NoStop}%
\bibitem [{\citenamefont {Spino}\ and\ \citenamefont
  {Matveev}(2022)}]{spino2022development}%
  \BibitemOpen
  \bibfield  {author} {\bibinfo {author} {\bibfnamefont {P.}~\bibnamefont
  {Spino}}\ and\ \bibinfo {author} {\bibfnamefont {K.~I.}\ \bibnamefont
  {Matveev}},\ }\bibfield  {title} {\bibinfo {title} {Development and testing
  of unmanned semi-submersible vehicle},\ }\href@noop {} {\bibfield  {journal}
  {\bibinfo  {journal} {Unmanned Systems}\ }\textbf {\bibinfo {volume} {11}}
  (\bibinfo {year} {2022})}\BibitemShut {NoStop}%
\bibitem [{\citenamefont {Vennell}\ \emph {et~al.}(2006)\citenamefont
  {Vennell}, \citenamefont {Pease},\ and\ \citenamefont
  {Wilson}}]{vennell2006wave}%
  \BibitemOpen
  \bibfield  {author} {\bibinfo {author} {\bibfnamefont {R.}~\bibnamefont
  {Vennell}}, \bibinfo {author} {\bibfnamefont {D.}~\bibnamefont {Pease}},\
  and\ \bibinfo {author} {\bibfnamefont {B.}~\bibnamefont {Wilson}},\
  }\bibfield  {title} {\bibinfo {title} {Wave drag on human swimmers},\
  }\href@noop {} {\bibfield  {journal} {\bibinfo  {journal} {Journal of
  Biomechanics}\ }\textbf {\bibinfo {volume} {39}} (\bibinfo {year}
  {2006})}\BibitemShut {NoStop}%
\bibitem [{\citenamefont {James}(2016)}]{james2016search}%
  \BibitemOpen
  \bibfield  {author} {\bibinfo {author} {\bibfnamefont {M.~C.}\ \bibnamefont
  {James}},\ }\emph {\bibinfo {title} {On the search to reduce a swimmer’s
  resistance: Surface-piercing bluff bodies over the critical Re-Fr range}},\
  \href@noop {} {Ph.D. thesis},\ \bibinfo  {school} {Universty of Southampton}
  (\bibinfo {year} {2016})\BibitemShut {NoStop}%
\bibitem [{\citenamefont {Jami}\ \emph {et~al.}(2021)\citenamefont {Jami},
  \citenamefont {Gustafson}, \citenamefont {Steinmann}, \citenamefont
  {Pi{\~n}eirua},\ and\ \citenamefont {Casas}}]{jami2021overcoming}%
  \BibitemOpen
  \bibfield  {author} {\bibinfo {author} {\bibfnamefont {L.}~\bibnamefont
  {Jami}}, \bibinfo {author} {\bibfnamefont {G.~T.}\ \bibnamefont {Gustafson}},
  \bibinfo {author} {\bibfnamefont {T.}~\bibnamefont {Steinmann}}, \bibinfo
  {author} {\bibfnamefont {M.}~\bibnamefont {Pi{\~n}eirua}},\ and\ \bibinfo
  {author} {\bibfnamefont {J.}~\bibnamefont {Casas}},\ }\bibfield  {title}
  {\bibinfo {title} {Overcoming drag at the water-air interface constrains body
  size in whirligig beetles},\ }\href@noop {} {\bibfield  {journal} {\bibinfo
  {journal} {Fluids}\ }\textbf {\bibinfo {volume} {6}} (\bibinfo {year}
  {2021})}\BibitemShut {NoStop}%
\bibitem [{\citenamefont {Hay}(1947)}]{hay1947}%
  \BibitemOpen
  \bibfield  {author} {\bibinfo {author} {\bibfnamefont {A.}~\bibnamefont
  {Hay}},\ }\bibfield  {title} {\bibinfo {title} {Flow about semi-submerged
  cylinders of finite length},\ }\href@noop {} {\bibfield  {journal} {\bibinfo
  {journal} {Princeton University}\ } (\bibinfo {year} {1947})}\BibitemShut
  {NoStop}%
\bibitem [{\citenamefont {Chaplin}\ and\ \citenamefont
  {Teigen}(2003)}]{chaplin2003steady}%
  \BibitemOpen
  \bibfield  {author} {\bibinfo {author} {\bibfnamefont {J.~R.}\ \bibnamefont
  {Chaplin}}\ and\ \bibinfo {author} {\bibfnamefont {P.}~\bibnamefont
  {Teigen}},\ }\bibfield  {title} {\bibinfo {title} {Steady flow past a
  vertical surface-piercing circular cylinder},\ }\href@noop {} {\bibfield
  {journal} {\bibinfo  {journal} {Journal of Fluids and Structures}\ }\textbf
  {\bibinfo {volume} {18}} (\bibinfo {year} {2003})}\BibitemShut {NoStop}%
\bibitem [{\citenamefont {Rapha\"el}\ and\ \citenamefont
  {de~Gennes}(1996)}]{raphael1996}%
  \BibitemOpen
  \bibfield  {author} {\bibinfo {author} {\bibfnamefont {E.}~\bibnamefont
  {Rapha\"el}}\ and\ \bibinfo {author} {\bibfnamefont {P.-G.}\ \bibnamefont
  {de~Gennes}},\ }\bibfield  {title} {\bibinfo {title} {Capillary gravity waves
  caused by a moving disturbance: Wave resistance},\ }\href
  {https://link.aps.org/doi/10.1103/PhysRevE.53.3448} {\bibfield  {journal}
  {\bibinfo  {journal} {Physical Review E}\ }\textbf {\bibinfo {volume} {53}}
  (\bibinfo {year} {1996})}\BibitemShut {NoStop}%
\bibitem [{\citenamefont {Browaeys}\ \emph {et~al.}(2001)\citenamefont
  {Browaeys}, \citenamefont {Bacri}, \citenamefont {Perzynski},\ and\
  \citenamefont {Shliomis}}]{browaeys2001capillary}%
  \BibitemOpen
  \bibfield  {author} {\bibinfo {author} {\bibfnamefont {J.}~\bibnamefont
  {Browaeys}}, \bibinfo {author} {\bibfnamefont {J.-C.}\ \bibnamefont {Bacri}},
  \bibinfo {author} {\bibfnamefont {R.}~\bibnamefont {Perzynski}},\ and\
  \bibinfo {author} {\bibfnamefont {M.}~\bibnamefont {Shliomis}},\ }\bibfield
  {title} {\bibinfo {title} {Capillary-gravity wave resistance in ordinary and
  magnetic fluids},\ }\href@noop {} {\bibfield  {journal} {\bibinfo  {journal}
  {Europhysics Letters}\ }\textbf {\bibinfo {volume} {53}} (\bibinfo {year}
  {2001})}\BibitemShut {NoStop}%
\bibitem [{\citenamefont {Burghelea}\ and\ \citenamefont
  {Steinberg}(2001)}]{burghelea2001}%
  \BibitemOpen
  \bibfield  {author} {\bibinfo {author} {\bibfnamefont {T.}~\bibnamefont
  {Burghelea}}\ and\ \bibinfo {author} {\bibfnamefont {V.}~\bibnamefont
  {Steinberg}},\ }\bibfield  {title} {\bibinfo {title} {Onset of wave drag due
  to generation of capillary-gravity waves by a moving object as a critical
  phenomenon},\ }\href {https://link.aps.org/doi/10.1103/PhysRevLett.86.2557}
  {\bibfield  {journal} {\bibinfo  {journal} {Physical Review Letters}\
  }\textbf {\bibinfo {volume} {86}} (\bibinfo {year} {2001})}\BibitemShut
  {NoStop}%
\bibitem [{\citenamefont {Burghelea}\ and\ \citenamefont
  {Steinberg}(2002)}]{burghelea2002}%
  \BibitemOpen
  \bibfield  {author} {\bibinfo {author} {\bibfnamefont {T.}~\bibnamefont
  {Burghelea}}\ and\ \bibinfo {author} {\bibfnamefont {V.}~\bibnamefont
  {Steinberg}},\ }\bibfield  {title} {\bibinfo {title} {Wave drag due to
  generation of capillary-gravity surface waves},\ }\href@noop {} {\bibfield
  {journal} {\bibinfo  {journal} {Physical Review E}\ }\textbf {\bibinfo
  {volume} {66}} (\bibinfo {year} {2002})}\BibitemShut {NoStop}%
\bibitem [{\citenamefont {Chevy}\ and\ \citenamefont
  {Rapha{\"e}l}(2003)}]{chevy2003capillary}%
  \BibitemOpen
  \bibfield  {author} {\bibinfo {author} {\bibfnamefont {F.}~\bibnamefont
  {Chevy}}\ and\ \bibinfo {author} {\bibfnamefont {E.}~\bibnamefont
  {Rapha{\"e}l}},\ }\bibfield  {title} {\bibinfo {title} {Capillary gravity
  waves: A “fixed-depth" analysis},\ }\href@noop {} {\bibfield  {journal}
  {\bibinfo  {journal} {Europhysics Letters}\ }\textbf {\bibinfo {volume} {61}}
  (\bibinfo {year} {2003})}\BibitemShut {NoStop}%
\bibitem [{\citenamefont {Richard}\ and\ \citenamefont
  {Raphael}(1999)}]{richard1999capillary}%
  \BibitemOpen
  \bibfield  {author} {\bibinfo {author} {\bibfnamefont {D.}~\bibnamefont
  {Richard}}\ and\ \bibinfo {author} {\bibfnamefont {E.}~\bibnamefont
  {Raphael}},\ }\bibfield  {title} {\bibinfo {title} {Capillary-gravity waves:
  The effect of viscosity on the wave resistance},\ }\href@noop {} {\bibfield
  {journal} {\bibinfo  {journal} {Europhysics Letters}\ }\textbf {\bibinfo
  {volume} {48}} (\bibinfo {year} {1999})}\BibitemShut {NoStop}%
\bibitem [{\citenamefont {Merrer}\ \emph {et~al.}(2011)\citenamefont {Merrer},
  \citenamefont {Clanet}, \citenamefont {Quéré}, \citenamefont {Élie
  Raphaël},\ and\ \citenamefont {Chevy}}]{merrer2011}%
  \BibitemOpen
  \bibfield  {author} {\bibinfo {author} {\bibfnamefont {M.~L.}\ \bibnamefont
  {Merrer}}, \bibinfo {author} {\bibfnamefont {C.}~\bibnamefont {Clanet}},
  \bibinfo {author} {\bibfnamefont {D.}~\bibnamefont {Quéré}}, \bibinfo
  {author} {\bibnamefont {Élie Raphaël}},\ and\ \bibinfo {author}
  {\bibfnamefont {F.}~\bibnamefont {Chevy}},\ }\bibfield  {title} {\bibinfo
  {title} {Wave drag on floating bodies},\ }\href@noop {} {\bibfield  {journal}
  {\bibinfo  {journal} {Proceedings of the National Academy of Sciences}\
  }\textbf {\bibinfo {volume} {108}} (\bibinfo {year} {2011})}\BibitemShut
  {NoStop}%
\bibitem [{\citenamefont {Denny}(1993)}]{denny1993air}%
  \BibitemOpen
  \bibfield  {author} {\bibinfo {author} {\bibfnamefont {M.}~\bibnamefont
  {Denny}},\ }\href@noop {} {\emph {\bibinfo {title} {Air and water: the
  biology and physics of life's media}}}\ (\bibinfo  {publisher} {Princeton
  University Press},\ \bibinfo {year} {1993})\BibitemShut {NoStop}%
\bibitem [{\citenamefont {Hu}\ \emph {et~al.}(2003)\citenamefont {Hu},
  \citenamefont {Chan},\ and\ \citenamefont {Bush}}]{hu2003hydrodynamics}%
  \BibitemOpen
  \bibfield  {author} {\bibinfo {author} {\bibfnamefont {D.~L.}\ \bibnamefont
  {Hu}}, \bibinfo {author} {\bibfnamefont {B.}~\bibnamefont {Chan}},\ and\
  \bibinfo {author} {\bibfnamefont {J.~W.}\ \bibnamefont {Bush}},\ }\bibfield
  {title} {\bibinfo {title} {The hydrodynamics of water strider locomotion},\
  }\href@noop {} {\bibfield  {journal} {\bibinfo  {journal} {Nature}\ }\textbf
  {\bibinfo {volume} {424}} (\bibinfo {year} {2003})}\BibitemShut {NoStop}%
\bibitem [{\citenamefont {Benusiglio}\ \emph {et~al.}(2015)\citenamefont
  {Benusiglio}, \citenamefont {Chevy}, \citenamefont {Rapha{\"e}l},\ and\
  \citenamefont {Clanet}}]{benusiglio2015wave}%
  \BibitemOpen
  \bibfield  {author} {\bibinfo {author} {\bibfnamefont {A.}~\bibnamefont
  {Benusiglio}}, \bibinfo {author} {\bibfnamefont {F.}~\bibnamefont {Chevy}},
  \bibinfo {author} {\bibfnamefont {{\'E}.}~\bibnamefont {Rapha{\"e}l}},\ and\
  \bibinfo {author} {\bibfnamefont {C.}~\bibnamefont {Clanet}},\ }\bibfield
  {title} {\bibinfo {title} {Wave drag on a submerged sphere},\ }\href@noop {}
  {\bibfield  {journal} {\bibinfo  {journal} {Physics of Fluids}\ }\textbf
  {\bibinfo {volume} {27}} (\bibinfo {year} {2015})}\BibitemShut {NoStop}%
\bibitem [{\citenamefont {Kamoliddinov}\ \emph {et~al.}(2021)\citenamefont
  {Kamoliddinov}, \citenamefont {Vakarelski},\ and\ \citenamefont
  {Thoroddsen}}]{kamoliddinov2021}%
  \BibitemOpen
  \bibfield  {author} {\bibinfo {author} {\bibfnamefont {F.}~\bibnamefont
  {Kamoliddinov}}, \bibinfo {author} {\bibfnamefont {I.~U.}\ \bibnamefont
  {Vakarelski}},\ and\ \bibinfo {author} {\bibfnamefont {S.~T.}\ \bibnamefont
  {Thoroddsen}},\ }\bibfield  {title} {\bibinfo {title} {{Hydrodynamic regimes
  and drag on horizontally pulled floating spheres}},\ }\href@noop {}
  {\bibfield  {journal} {\bibinfo  {journal} {Physics of Fluids}\ }\textbf
  {\bibinfo {volume} {33}} (\bibinfo {year} {2021})},\ \bibinfo {note}
  {093308}\BibitemShut {NoStop}%
\bibitem [{\citenamefont {James}\ \emph {et~al.}(2015)\citenamefont {James},
  \citenamefont {Forrester}, \citenamefont {Hudson}, \citenamefont {Taunton},\
  and\ \citenamefont {Turnock}}]{james2015}%
  \BibitemOpen
  \bibfield  {author} {\bibinfo {author} {\bibfnamefont {M.}~\bibnamefont
  {James}}, \bibinfo {author} {\bibfnamefont {A.}~\bibnamefont {Forrester}},
  \bibinfo {author} {\bibfnamefont {D.}~\bibnamefont {Hudson}}, \bibinfo
  {author} {\bibfnamefont {D.}~\bibnamefont {Taunton}},\ and\ \bibinfo {author}
  {\bibfnamefont {S.}~\bibnamefont {Turnock}},\ }\bibfield  {title} {\bibinfo
  {title} {Experimental study of the transitional flow of a sphere located at
  the free surface},\ }in\ \href {https://eprints.soton.ac.uk/382010/} {\emph
  {\bibinfo {booktitle} {9th International Workshop on Ship and Marine
  Hydrodynamics}}}\ (\bibinfo {year} {2015})\BibitemShut {NoStop}%
\bibitem [{\citenamefont {Morrison}(2013)}]{morrison2013introduction}%
  \BibitemOpen
  \bibfield  {author} {\bibinfo {author} {\bibfnamefont {F.~A.}\ \bibnamefont
  {Morrison}},\ }\href@noop {} {\emph {\bibinfo {title} {An introduction to
  fluid mechanics}}}\ (\bibinfo  {publisher} {Cambridge University Press},\
  \bibinfo {year} {2013})\BibitemShut {NoStop}%
\bibitem [{\citenamefont {Bell}\ and\ \citenamefont
  {Mehta}(1988)}]{bell1988contraction}%
  \BibitemOpen
  \bibfield  {author} {\bibinfo {author} {\bibfnamefont {J.~H.}\ \bibnamefont
  {Bell}}\ and\ \bibinfo {author} {\bibfnamefont {R.~D.}\ \bibnamefont
  {Mehta}},\ }\bibfield  {title} {\bibinfo {title} {Contraction design for
  small low-speed wind tunnels},\ }\href@noop {} {\bibfield  {journal}
  {\bibinfo  {journal} {NASA Technical Report}\ } (\bibinfo {year}
  {1988})}\BibitemShut {NoStop}%
\bibitem [{\citenamefont {Thielicke}\ and\ \citenamefont
  {Sonntag}(2021)}]{thielicke2021}%
  \BibitemOpen
  \bibfield  {author} {\bibinfo {author} {\bibfnamefont {W.}~\bibnamefont
  {Thielicke}}\ and\ \bibinfo {author} {\bibfnamefont {R.}~\bibnamefont
  {Sonntag}},\ }\bibfield  {title} {\bibinfo {title} {Particle image
  velocimetry for {MATLAB}: Accuracy and enhanced algorithms in {PIVlab}},\
  }\href {https://doi.org/10.5334/jors.334} {\bibfield  {journal} {\bibinfo
  {journal} {Journal of Open Research Software}\ }\textbf {\bibinfo {volume}
  {9}} (\bibinfo {year} {2021})}\BibitemShut {NoStop}%
\bibitem [{\citenamefont {Galeano-Rios}\ \emph {et~al.}(2021)\citenamefont
  {Galeano-Rios}, \citenamefont {Cimpeanu}, \citenamefont {Bauman},
  \citenamefont {MacEwen}, \citenamefont {Milewski},\ and\ \citenamefont
  {Harris}}]{galeano2021capillary}%
  \BibitemOpen
  \bibfield  {author} {\bibinfo {author} {\bibfnamefont {C.~A.}\ \bibnamefont
  {Galeano-Rios}}, \bibinfo {author} {\bibfnamefont {R.}~\bibnamefont
  {Cimpeanu}}, \bibinfo {author} {\bibfnamefont {I.~A.}\ \bibnamefont
  {Bauman}}, \bibinfo {author} {\bibfnamefont {A.}~\bibnamefont {MacEwen}},
  \bibinfo {author} {\bibfnamefont {P.~A.}\ \bibnamefont {Milewski}},\ and\
  \bibinfo {author} {\bibfnamefont {D.~M.}\ \bibnamefont {Harris}},\ }\bibfield
   {title} {\bibinfo {title} {Capillary-scale solid rebounds: experiments,
  modelling and simulations},\ }\href@noop {} {\bibfield  {journal} {\bibinfo
  {journal} {Journal of Fluid Mechanics}\ }\textbf {\bibinfo {volume} {912}},\
  \bibinfo {pages} {A17} (\bibinfo {year} {2021})}\BibitemShut {NoStop}%
\bibitem [{\citenamefont {Ed-Daoui}\ \emph {et~al.}(2019)\citenamefont
  {Ed-Daoui}, \citenamefont {Benelmostafa},\ and\ \citenamefont
  {Dahmani}}]{EDDAOUI2019746}%
  \BibitemOpen
  \bibfield  {author} {\bibinfo {author} {\bibfnamefont {A.}~\bibnamefont
  {Ed-Daoui}}, \bibinfo {author} {\bibfnamefont {M.}~\bibnamefont
  {Benelmostafa}},\ and\ \bibinfo {author} {\bibfnamefont {M.}~\bibnamefont
  {Dahmani}},\ }\bibfield  {title} {\bibinfo {title} {Study of the viscoelastic
  properties of the agarose gel},\ }\href
  {https://doi.org/https://doi.org/10.1016/j.matpr.2019.04.036} {\bibfield
  {journal} {\bibinfo  {journal} {Materials Today: Proceedings}\ }\textbf
  {\bibinfo {volume} {13}},\ \bibinfo {pages} {746} (\bibinfo {year}
  {2019})}\BibitemShut {NoStop}%
\bibitem [{\citenamefont {Andersson}\ \emph {et~al.}(2017)\citenamefont
  {Andersson}, \citenamefont {Str{\"o}m}, \citenamefont {Geb{\"a}ck},\ and\
  \citenamefont {Larsson}}]{andersson2017dynamics}%
  \BibitemOpen
  \bibfield  {author} {\bibinfo {author} {\bibfnamefont {J.}~\bibnamefont
  {Andersson}}, \bibinfo {author} {\bibfnamefont {A.}~\bibnamefont
  {Str{\"o}m}}, \bibinfo {author} {\bibfnamefont {T.}~\bibnamefont
  {Geb{\"a}ck}},\ and\ \bibinfo {author} {\bibfnamefont {A.}~\bibnamefont
  {Larsson}},\ }\bibfield  {title} {\bibinfo {title} {Dynamics of capillary
  transport in semi-solid channels},\ }\href@noop {} {\bibfield  {journal}
  {\bibinfo  {journal} {Soft Matter}\ }\textbf {\bibinfo {volume} {13}}
  (\bibinfo {year} {2017})}\BibitemShut {NoStop}%
\bibitem [{\citenamefont {Osti}\ \emph {et~al.}(2009)\citenamefont {Osti},
  \citenamefont {Wolf},\ and\ \citenamefont {Philippi}}]{osti2009}%
  \BibitemOpen
  \bibfield  {author} {\bibinfo {author} {\bibfnamefont {G.~B.~F.}\
  \bibnamefont {Osti}}, \bibinfo {author} {\bibfnamefont {F.~G.}\ \bibnamefont
  {Wolf}},\ and\ \bibinfo {author} {\bibfnamefont {P.~C.}\ \bibnamefont
  {Philippi}},\ }\bibfield  {title} {\bibinfo {title} {Spreading of liquid
  drops on acrylic surfaces},\ }in\ \href
  {https://www.abcm.org.br/anais/cobem/2009/pdf/COB09-1947.pdf} {\emph
  {\bibinfo {booktitle} {20th International Congress of Mechanical
  Engineering}}}\ (\bibinfo {year} {2009})\BibitemShut {NoStop}%
\bibitem [{\citenamefont {Weisensee}\ \emph {et~al.}(2017)\citenamefont
  {Weisensee}, \citenamefont {Ma}, \citenamefont {Shin}, \citenamefont {Tian},
  \citenamefont {Chang}, \citenamefont {King},\ and\ \citenamefont
  {Miljkovic}}]{weisensee2017droplet}%
  \BibitemOpen
  \bibfield  {author} {\bibinfo {author} {\bibfnamefont {P.~B.}\ \bibnamefont
  {Weisensee}}, \bibinfo {author} {\bibfnamefont {J.}~\bibnamefont {Ma}},
  \bibinfo {author} {\bibfnamefont {Y.~H.}\ \bibnamefont {Shin}}, \bibinfo
  {author} {\bibfnamefont {J.}~\bibnamefont {Tian}}, \bibinfo {author}
  {\bibfnamefont {Y.}~\bibnamefont {Chang}}, \bibinfo {author} {\bibfnamefont
  {W.~P.}\ \bibnamefont {King}},\ and\ \bibinfo {author} {\bibfnamefont
  {N.}~\bibnamefont {Miljkovic}},\ }\bibfield  {title} {\bibinfo {title}
  {Droplet impact on vibrating superhydrophobic surfaces},\ }\href@noop {}
  {\bibfield  {journal} {\bibinfo  {journal} {Physical Review Fluids}\ }\textbf
  {\bibinfo {volume} {2}} (\bibinfo {year} {2017})}\BibitemShut {NoStop}%
\bibitem [{\citenamefont {Bazilevs}\ \emph
  {et~al.}(2007{\natexlab{a}})\citenamefont {Bazilevs}, \citenamefont {Calo},
  \citenamefont {Cottrell}, \citenamefont {Hughes}, \citenamefont {Reali},\
  and\ \citenamefont {Scovazzi}}]{bazilevs2007variational}%
  \BibitemOpen
  \bibfield  {author} {\bibinfo {author} {\bibfnamefont {Y.}~\bibnamefont
  {Bazilevs}}, \bibinfo {author} {\bibfnamefont {V.}~\bibnamefont {Calo}},
  \bibinfo {author} {\bibfnamefont {J.}~\bibnamefont {Cottrell}}, \bibinfo
  {author} {\bibfnamefont {T.}~\bibnamefont {Hughes}}, \bibinfo {author}
  {\bibfnamefont {A.}~\bibnamefont {Reali}},\ and\ \bibinfo {author}
  {\bibfnamefont {G.}~\bibnamefont {Scovazzi}},\ }\bibfield  {title} {\bibinfo
  {title} {Variational multiscale residual-based turbulence modeling for large
  eddy simulation of incompressible flows},\ }\href@noop {} {\bibfield
  {journal} {\bibinfo  {journal} {Computer methods in applied mechanics and
  engineering}\ }\textbf {\bibinfo {volume} {197}},\ \bibinfo {pages} {173}
  (\bibinfo {year} {2007}{\natexlab{a}})}\BibitemShut {NoStop}%
\bibitem [{\citenamefont {Hsu}\ \emph {et~al.}(2012)\citenamefont {Hsu},
  \citenamefont {Akkerman},\ and\ \citenamefont {Bazilevs}}]{hsu2012wind}%
  \BibitemOpen
  \bibfield  {author} {\bibinfo {author} {\bibfnamefont {M.-C.}\ \bibnamefont
  {Hsu}}, \bibinfo {author} {\bibfnamefont {I.}~\bibnamefont {Akkerman}},\ and\
  \bibinfo {author} {\bibfnamefont {Y.}~\bibnamefont {Bazilevs}},\ }\bibfield
  {title} {\bibinfo {title} {Wind turbine aerodynamics using {ALE}--{VMS}:
  Validation and the role of weakly enforced boundary conditions},\ }\href@noop
  {} {\bibfield  {journal} {\bibinfo  {journal} {Computational Mechanics}\
  }\textbf {\bibinfo {volume} {50}},\ \bibinfo {pages} {499} (\bibinfo {year}
  {2012})}\BibitemShut {NoStop}%
\bibitem [{\citenamefont {Hsu}\ \emph {et~al.}(2014)\citenamefont {Hsu},
  \citenamefont {Akkerman},\ and\ \citenamefont {Bazilevs}}]{hsu2014finite}%
  \BibitemOpen
  \bibfield  {author} {\bibinfo {author} {\bibfnamefont {M.-C.}\ \bibnamefont
  {Hsu}}, \bibinfo {author} {\bibfnamefont {I.}~\bibnamefont {Akkerman}},\ and\
  \bibinfo {author} {\bibfnamefont {Y.}~\bibnamefont {Bazilevs}},\ }\bibfield
  {title} {\bibinfo {title} {Finite element simulation of wind turbine
  aerodynamics: validation study using {NREL} {Phase} {VI} experiment},\
  }\href@noop {} {\bibfield  {journal} {\bibinfo  {journal} {Wind Energy}\
  }\textbf {\bibinfo {volume} {17}},\ \bibinfo {pages} {461} (\bibinfo {year}
  {2014})}\BibitemShut {NoStop}%
\bibitem [{\citenamefont {Bazilevs}\ \emph {et~al.}(2014)\citenamefont
  {Bazilevs}, \citenamefont {Yan}, \citenamefont {De~Stadler},\ and\
  \citenamefont {Sarkar}}]{bazilevs2014computation}%
  \BibitemOpen
  \bibfield  {author} {\bibinfo {author} {\bibfnamefont {Y.}~\bibnamefont
  {Bazilevs}}, \bibinfo {author} {\bibfnamefont {J.}~\bibnamefont {Yan}},
  \bibinfo {author} {\bibfnamefont {M.}~\bibnamefont {De~Stadler}},\ and\
  \bibinfo {author} {\bibfnamefont {S.}~\bibnamefont {Sarkar}},\ }\bibfield
  {title} {\bibinfo {title} {Computation of the flow over a sphere at re= 3700:
  A comparison of uniform and turbulent inflow conditions},\ }\href@noop {}
  {\bibfield  {journal} {\bibinfo  {journal} {Journal of Applied Mechanics}\
  }\textbf {\bibinfo {volume} {81}},\ \bibinfo {pages} {121003} (\bibinfo
  {year} {2014})}\BibitemShut {NoStop}%
\bibitem [{\citenamefont {Pal}\ and\ \citenamefont
  {Gohari}(2015)}]{bazilevs2015ale}%
  \BibitemOpen
  \bibfield  {author} {\bibinfo {author} {\bibfnamefont {A.}~\bibnamefont
  {Pal}}\ and\ \bibinfo {author} {\bibfnamefont {S.}~\bibnamefont {Gohari}},\
  }\bibfield  {title} {\bibinfo {title} {{ALE--VMS} formulation for stratified
  turbulent incompressible flows with applications},\ }\href@noop {} {\bibfield
   {journal} {\bibinfo  {journal} {Mathematical Models and Methods in Applied
  Sciences}\ }\textbf {\bibinfo {volume} {25}},\ \bibinfo {pages} {2349}
  (\bibinfo {year} {2015})}\BibitemShut {NoStop}%
\bibitem [{\citenamefont {Xu}\ \emph {et~al.}(2016)\citenamefont {Xu},
  \citenamefont {Schillinger}, \citenamefont {Kamensky}, \citenamefont
  {Varduhn}, \citenamefont {Wang},\ and\ \citenamefont
  {Hsu}}]{xu2016tetrahedral}%
  \BibitemOpen
  \bibfield  {author} {\bibinfo {author} {\bibfnamefont {F.}~\bibnamefont
  {Xu}}, \bibinfo {author} {\bibfnamefont {D.}~\bibnamefont {Schillinger}},
  \bibinfo {author} {\bibfnamefont {D.}~\bibnamefont {Kamensky}}, \bibinfo
  {author} {\bibfnamefont {V.}~\bibnamefont {Varduhn}}, \bibinfo {author}
  {\bibfnamefont {C.}~\bibnamefont {Wang}},\ and\ \bibinfo {author}
  {\bibfnamefont {M.-C.}\ \bibnamefont {Hsu}},\ }\bibfield  {title} {\bibinfo
  {title} {The tetrahedral finite cell method for fluids: Immersogeometric
  analysis of turbulent flow around complex geometries},\ }\href@noop {}
  {\bibfield  {journal} {\bibinfo  {journal} {Computers \& Fluids}\ }\textbf
  {\bibinfo {volume} {141}},\ \bibinfo {pages} {135} (\bibinfo {year}
  {2016})}\BibitemShut {NoStop}%
\bibitem [{\citenamefont {Yan}\ \emph {et~al.}(2017{\natexlab{a}})\citenamefont
  {Yan}, \citenamefont {Korobenko}, \citenamefont {Tejada-Martinez},\ and\
  \citenamefont {Golshan}}]{yan2017new}%
  \BibitemOpen
  \bibfield  {author} {\bibinfo {author} {\bibfnamefont {J.}~\bibnamefont
  {Yan}}, \bibinfo {author} {\bibfnamefont {A.}~\bibnamefont {Korobenko}},
  \bibinfo {author} {\bibfnamefont {A.}~\bibnamefont {Tejada-Martinez}},\ and\
  \bibinfo {author} {\bibfnamefont {R.}~\bibnamefont {Golshan}},\ }\bibfield
  {title} {\bibinfo {title} {A new variational multiscale formulation for
  stratified incompressible turbulent flows},\ }\href@noop {} {\bibfield
  {journal} {\bibinfo  {journal} {Computers \& Fluids}\ }\textbf {\bibinfo
  {volume} {158}},\ \bibinfo {pages} {150} (\bibinfo {year}
  {2017}{\natexlab{a}})}\BibitemShut {NoStop}%
\bibitem [{\citenamefont {Zhao}\ and\ \citenamefont
  {Yan}(2022)}]{zhao2022enriched}%
  \BibitemOpen
  \bibfield  {author} {\bibinfo {author} {\bibfnamefont {Z.}~\bibnamefont
  {Zhao}}\ and\ \bibinfo {author} {\bibfnamefont {J.}~\bibnamefont {Yan}},\
  }\bibfield  {title} {\bibinfo {title} {Enriched immersed boundary method
  ({EIBM}) for interface-coupled multi-physics and applications to convective
  conjugate heat transfer},\ }\href@noop {} {\bibfield  {journal} {\bibinfo
  {journal} {Computer Methods in Applied Mechanics and Engineering}\ }\textbf
  {\bibinfo {volume} {401}},\ \bibinfo {pages} {115667} (\bibinfo {year}
  {2022})}\BibitemShut {NoStop}%
\bibitem [{\citenamefont {Hsu}\ \emph {et~al.}(2011)\citenamefont {Hsu},
  \citenamefont {Akkerman},\ and\ \citenamefont {Bazilevs}}]{hsu2011high}%
  \BibitemOpen
  \bibfield  {author} {\bibinfo {author} {\bibfnamefont {M.-C.}\ \bibnamefont
  {Hsu}}, \bibinfo {author} {\bibfnamefont {I.}~\bibnamefont {Akkerman}},\ and\
  \bibinfo {author} {\bibfnamefont {Y.}~\bibnamefont {Bazilevs}},\ }\bibfield
  {title} {\bibinfo {title} {High-performance computing of wind turbine
  aerodynamics using isogeometric analysis},\ }\href@noop {} {\bibfield
  {journal} {\bibinfo  {journal} {Computers \& Fluids}\ }\textbf {\bibinfo
  {volume} {49}},\ \bibinfo {pages} {93} (\bibinfo {year} {2011})}\BibitemShut
  {NoStop}%
\bibitem [{\citenamefont {van Opstal}\ \emph {et~al.}(2017)\citenamefont {van
  Opstal}, \citenamefont {Yan}, \citenamefont {Coley}, \citenamefont {Evans},
  \citenamefont {Kvamsdal},\ and\ \citenamefont
  {Bazilevs}}]{van2017isogeometric}%
  \BibitemOpen
  \bibfield  {author} {\bibinfo {author} {\bibfnamefont {T.~M.}\ \bibnamefont
  {van Opstal}}, \bibinfo {author} {\bibfnamefont {J.}~\bibnamefont {Yan}},
  \bibinfo {author} {\bibfnamefont {C.}~\bibnamefont {Coley}}, \bibinfo
  {author} {\bibfnamefont {J.~A.}\ \bibnamefont {Evans}}, \bibinfo {author}
  {\bibfnamefont {T.}~\bibnamefont {Kvamsdal}},\ and\ \bibinfo {author}
  {\bibfnamefont {Y.}~\bibnamefont {Bazilevs}},\ }\bibfield  {title} {\bibinfo
  {title} {Isogeometric divergence-conforming variational multiscale
  formulation of incompressible turbulent flows},\ }\href@noop {} {\bibfield
  {journal} {\bibinfo  {journal} {Computer Methods in Applied Mechanics and
  Engineering}\ }\textbf {\bibinfo {volume} {316}},\ \bibinfo {pages} {859}
  (\bibinfo {year} {2017})}\BibitemShut {NoStop}%
\bibitem [{\citenamefont {Bazilevs}\ \emph {et~al.}(2013)\citenamefont
  {Bazilevs}, \citenamefont {Akkerman},\ and\ \citenamefont
  {Benson}}]{bazilevs2013isogeometric}%
  \BibitemOpen
  \bibfield  {author} {\bibinfo {author} {\bibfnamefont {Y.}~\bibnamefont
  {Bazilevs}}, \bibinfo {author} {\bibfnamefont {I.}~\bibnamefont {Akkerman}},\
  and\ \bibinfo {author} {\bibfnamefont {D.}~\bibnamefont {Benson}},\
  }\bibfield  {title} {\bibinfo {title} {Isogeometric analysis of {Lagrangian}
  hydrodynamics},\ }\href@noop {} {\bibfield  {journal} {\bibinfo  {journal}
  {Journal of Computational Physics}\ }\textbf {\bibinfo {volume} {243}},\
  \bibinfo {pages} {224} (\bibinfo {year} {2013})}\BibitemShut {NoStop}%
\bibitem [{\citenamefont {Bazilevs}\ \emph
  {et~al.}(2007{\natexlab{b}})\citenamefont {Bazilevs}, \citenamefont
  {Michler},\ and\ \citenamefont {Calo}}]{bazilevs2007weak}%
  \BibitemOpen
  \bibfield  {author} {\bibinfo {author} {\bibfnamefont {Y.}~\bibnamefont
  {Bazilevs}}, \bibinfo {author} {\bibfnamefont {C.}~\bibnamefont {Michler}},\
  and\ \bibinfo {author} {\bibfnamefont {V.}~\bibnamefont {Calo}},\ }\bibfield
  {title} {\bibinfo {title} {Weak {D}irichlet boundary conditions for
  wall-bounded turbulent flows},\ }\href@noop {} {\bibfield  {journal}
  {\bibinfo  {journal} {Computer Methods in Applied Mechanics and Engineering}\
  }\textbf {\bibinfo {volume} {196}},\ \bibinfo {pages} {4853} (\bibinfo {year}
  {2007}{\natexlab{b}})}\BibitemShut {NoStop}%
\bibitem [{\citenamefont {Bazilevs}\ and\ \citenamefont
  {Hughes}(2007)}]{Bazilevs07c}%
  \BibitemOpen
  \bibfield  {author} {\bibinfo {author} {\bibfnamefont {Y.}~\bibnamefont
  {Bazilevs}}\ and\ \bibinfo {author} {\bibfnamefont {T.~J.~R.}\ \bibnamefont
  {Hughes}},\ }\bibfield  {title} {\bibinfo {title} {Weak imposition of
  {D}irichlet boundary conditions in fluid mechanics},\ }\href@noop {}
  {\bibfield  {journal} {\bibinfo  {journal} {Computers \& Fluids}\ }\textbf
  {\bibinfo {volume} {36}},\ \bibinfo {pages} {12} (\bibinfo {year}
  {2007})}\BibitemShut {NoStop}%
\bibitem [{\citenamefont {Sussman}\ \emph {et~al.}(1994)\citenamefont
  {Sussman}, \citenamefont {Smereka},\ and\ \citenamefont
  {Osher}}]{sussman1994level}%
  \BibitemOpen
  \bibfield  {author} {\bibinfo {author} {\bibfnamefont {M.}~\bibnamefont
  {Sussman}}, \bibinfo {author} {\bibfnamefont {P.}~\bibnamefont {Smereka}},\
  and\ \bibinfo {author} {\bibfnamefont {S.}~\bibnamefont {Osher}},\ }\bibfield
   {title} {\bibinfo {title} {A level set approach for computing solutions to
  incompressible two-phase flow},\ }\href@noop {} {\bibfield  {journal}
  {\bibinfo  {journal} {Journal of Computational physics}\ }\textbf {\bibinfo
  {volume} {114}},\ \bibinfo {pages} {146} (\bibinfo {year}
  {1994})}\BibitemShut {NoStop}%
\bibitem [{\citenamefont {Osher}\ and\ \citenamefont
  {Sethian}(1988)}]{osher1988fronts}%
  \BibitemOpen
  \bibfield  {author} {\bibinfo {author} {\bibfnamefont {S.}~\bibnamefont
  {Osher}}\ and\ \bibinfo {author} {\bibfnamefont {J.~A.}\ \bibnamefont
  {Sethian}},\ }\bibfield  {title} {\bibinfo {title} {Fronts propagating with
  curvature-dependent speed: algorithms based on {H}amilton-{J}acobi
  formulations},\ }\href@noop {} {\bibfield  {journal} {\bibinfo  {journal}
  {Journal of Computational Physics}\ }\textbf {\bibinfo {volume} {79}},\
  \bibinfo {pages} {12} (\bibinfo {year} {1988})}\BibitemShut {NoStop}%
\bibitem [{\citenamefont {Zhu}\ and\ \citenamefont {Yan}(2021)}]{zhu2021mixed}%
  \BibitemOpen
  \bibfield  {author} {\bibinfo {author} {\bibfnamefont {Q.}~\bibnamefont
  {Zhu}}\ and\ \bibinfo {author} {\bibfnamefont {J.}~\bibnamefont {Yan}},\
  }\bibfield  {title} {\bibinfo {title} {A mixed
  interface-capturing/interface-tracking formulation for thermal multi-phase
  flows with emphasis on metal additive manufacturing processes},\ }\href@noop
  {} {\bibfield  {journal} {\bibinfo  {journal} {Computer Methods in Applied
  Mechanics and Engineering}\ }\textbf {\bibinfo {volume} {383}},\ \bibinfo
  {pages} {113910} (\bibinfo {year} {2021})}\BibitemShut {NoStop}%
\bibitem [{\citenamefont {Brackbill}\ \emph {et~al.}(1992)\citenamefont
  {Brackbill}, \citenamefont {Kothe},\ and\ \citenamefont
  {Zemach}}]{brackbill1992continuum}%
  \BibitemOpen
  \bibfield  {author} {\bibinfo {author} {\bibfnamefont {J.~U.}\ \bibnamefont
  {Brackbill}}, \bibinfo {author} {\bibfnamefont {D.~B.}\ \bibnamefont
  {Kothe}},\ and\ \bibinfo {author} {\bibfnamefont {C.}~\bibnamefont
  {Zemach}},\ }\bibfield  {title} {\bibinfo {title} {A continuum method for
  modeling surface tension},\ }\href@noop {} {\bibfield  {journal} {\bibinfo
  {journal} {Journal of computational physics}\ }\textbf {\bibinfo {volume}
  {100}},\ \bibinfo {pages} {335} (\bibinfo {year} {1992})}\BibitemShut
  {NoStop}%
\bibitem [{\citenamefont {Zhao}\ and\ \citenamefont
  {Yan}(2021)}]{zhao2021variational}%
  \BibitemOpen
  \bibfield  {author} {\bibinfo {author} {\bibfnamefont {Z.}~\bibnamefont
  {Zhao}}\ and\ \bibinfo {author} {\bibfnamefont {J.}~\bibnamefont {Yan}},\
  }\bibfield  {title} {\bibinfo {title} {Variational multi-scale modeling of
  interfacial flows with a balanced-force surface tension model},\ }\href@noop
  {} {\bibfield  {journal} {\bibinfo  {journal} {Mechanics Research
  Communications}\ }\textbf {\bibinfo {volume} {112}},\ \bibinfo {pages}
  {103608} (\bibinfo {year} {2021})}\BibitemShut {NoStop}%
\bibitem [{\citenamefont {Akkerman}\ and\ \citenamefont
  {Bazilevs}(2011)}]{akkerman2011isogeometric}%
  \BibitemOpen
  \bibfield  {author} {\bibinfo {author} {\bibfnamefont {I.}~\bibnamefont
  {Akkerman}}\ and\ \bibinfo {author} {\bibfnamefont {Y.}~\bibnamefont
  {Bazilevs}},\ }\bibfield  {title} {\bibinfo {title} {Isogeometric analysis of
  free-surface flow},\ }\href@noop {} {\bibfield  {journal} {\bibinfo
  {journal} {Journal of Computational Physics}\ }\textbf {\bibinfo {volume}
  {230}},\ \bibinfo {pages} {4137} (\bibinfo {year} {2011})}\BibitemShut
  {NoStop}%
\bibitem [{\citenamefont {Yan}\ \emph {et~al.}(2016)\citenamefont {Yan},
  \citenamefont {Korobenko}, \citenamefont {Deng},\ and\ \citenamefont
  {Bazilevs}}]{yan2016computational}%
  \BibitemOpen
  \bibfield  {author} {\bibinfo {author} {\bibfnamefont {J.}~\bibnamefont
  {Yan}}, \bibinfo {author} {\bibfnamefont {A.}~\bibnamefont {Korobenko}},
  \bibinfo {author} {\bibfnamefont {X.}~\bibnamefont {Deng}},\ and\ \bibinfo
  {author} {\bibfnamefont {Y.}~\bibnamefont {Bazilevs}},\ }\bibfield  {title}
  {\bibinfo {title} {Computational free-surface fluid--structure interaction
  with application to floating offshore wind turbines},\ }\href@noop {}
  {\bibfield  {journal} {\bibinfo  {journal} {Computers \& Fluids}\ }\textbf
  {\bibinfo {volume} {141}},\ \bibinfo {pages} {155} (\bibinfo {year}
  {2016})}\BibitemShut {NoStop}%
\bibitem [{\citenamefont {Yan}\ \emph {et~al.}(2017{\natexlab{b}})\citenamefont
  {Yan}, \citenamefont {Deng}, \citenamefont {Korobenko},\ and\ \citenamefont
  {Bazilevs}}]{yan2017free}%
  \BibitemOpen
  \bibfield  {author} {\bibinfo {author} {\bibfnamefont {J.}~\bibnamefont
  {Yan}}, \bibinfo {author} {\bibfnamefont {X.}~\bibnamefont {Deng}}, \bibinfo
  {author} {\bibfnamefont {A.}~\bibnamefont {Korobenko}},\ and\ \bibinfo
  {author} {\bibfnamefont {Y.}~\bibnamefont {Bazilevs}},\ }\bibfield  {title}
  {\bibinfo {title} {Free-surface flow modeling and simulation of
  horizontal-axis tidal-stream turbines},\ }\href@noop {} {\bibfield  {journal}
  {\bibinfo  {journal} {Computers \& Fluids}\ }\textbf {\bibinfo {volume}
  {158}},\ \bibinfo {pages} {157} (\bibinfo {year}
  {2017}{\natexlab{b}})}\BibitemShut {NoStop}%
\bibitem [{\citenamefont {Yan}\ \emph {et~al.}(2018)\citenamefont {Yan},
  \citenamefont {Yan}, \citenamefont {Lin},\ and\ \citenamefont
  {Wagner}}]{yan2018fully}%
  \BibitemOpen
  \bibfield  {author} {\bibinfo {author} {\bibfnamefont {J.}~\bibnamefont
  {Yan}}, \bibinfo {author} {\bibfnamefont {W.}~\bibnamefont {Yan}}, \bibinfo
  {author} {\bibfnamefont {S.}~\bibnamefont {Lin}},\ and\ \bibinfo {author}
  {\bibfnamefont {G.}~\bibnamefont {Wagner}},\ }\bibfield  {title} {\bibinfo
  {title} {A fully coupled finite element formulation for liquid--solid--gas
  thermo-fluid flow with melting and solidification},\ }\href@noop {}
  {\bibfield  {journal} {\bibinfo  {journal} {Computer Methods in Applied
  Mechanics and Engineering}\ }\textbf {\bibinfo {volume} {336}},\ \bibinfo
  {pages} {444} (\bibinfo {year} {2018})}\BibitemShut {NoStop}%
\bibitem [{\citenamefont {Yan}\ \emph {et~al.}(2019)\citenamefont {Yan},
  \citenamefont {Lin},\ and\ \citenamefont {Bazilevs}}]{yan2019isogeometric}%
  \BibitemOpen
  \bibfield  {author} {\bibinfo {author} {\bibfnamefont {J.}~\bibnamefont
  {Yan}}, \bibinfo {author} {\bibfnamefont {S.}~\bibnamefont {Lin}},\ and\
  \bibinfo {author} {\bibfnamefont {Y.}~\bibnamefont {Bazilevs}},\ }\bibfield
  {title} {\bibinfo {title} {Isogeometric analysis of multi-phase flows with
  surface tension and with application to dynamics of rising bubbles},\
  }\href@noop {} {\bibfield  {journal} {\bibinfo  {journal} {Computers \&
  Fluids}\ }\textbf {\bibinfo {volume} {179}},\ \bibinfo {pages} {777}
  (\bibinfo {year} {2019})}\BibitemShut {NoStop}%
\bibitem [{\citenamefont {Zhu}\ \emph {et~al.}(2020)\citenamefont {Zhu},
  \citenamefont {Xu}, \citenamefont {Xu}, \citenamefont {Hsu},\ and\
  \citenamefont {Yan}}]{zhu2020immersogeometric}%
  \BibitemOpen
  \bibfield  {author} {\bibinfo {author} {\bibfnamefont {Q.}~\bibnamefont
  {Zhu}}, \bibinfo {author} {\bibfnamefont {F.}~\bibnamefont {Xu}}, \bibinfo
  {author} {\bibfnamefont {S.}~\bibnamefont {Xu}}, \bibinfo {author}
  {\bibfnamefont {M.-C.}\ \bibnamefont {Hsu}},\ and\ \bibinfo {author}
  {\bibfnamefont {J.}~\bibnamefont {Yan}},\ }\bibfield  {title} {\bibinfo
  {title} {An immersogeometric formulation for free-surface flows with
  application to marine engineering problems},\ }\href@noop {} {\bibfield
  {journal} {\bibinfo  {journal} {Computer Methods in Applied Mechanics and
  Engineering}\ }\textbf {\bibinfo {volume} {361}},\ \bibinfo {pages} {112748}
  (\bibinfo {year} {2020})}\BibitemShut {NoStop}%
\bibitem [{\citenamefont {Yuan}\ and\ \citenamefont {Cho}(2012)}]{yuan2012bio}%
  \BibitemOpen
  \bibfield  {author} {\bibinfo {author} {\bibfnamefont {J.}~\bibnamefont
  {Yuan}}\ and\ \bibinfo {author} {\bibfnamefont {S.~K.}\ \bibnamefont {Cho}},\
  }\bibfield  {title} {\bibinfo {title} {Bio-inspired micro/mini propulsion at
  air-water interface: a review},\ }\href@noop {} {\bibfield  {journal}
  {\bibinfo  {journal} {Journal of Mechanical Science and Technology}\ }\textbf
  {\bibinfo {volume} {26}} (\bibinfo {year} {2012})}\BibitemShut {NoStop}%
\bibitem [{\citenamefont {Chen}\ \emph {et~al.}(2018)\citenamefont {Chen},
  \citenamefont {Doshi}, \citenamefont {Goldberg}, \citenamefont {Wang},\ and\
  \citenamefont {Wood}}]{chen2018controllable}%
  \BibitemOpen
  \bibfield  {author} {\bibinfo {author} {\bibfnamefont {Y.}~\bibnamefont
  {Chen}}, \bibinfo {author} {\bibfnamefont {N.}~\bibnamefont {Doshi}},
  \bibinfo {author} {\bibfnamefont {B.}~\bibnamefont {Goldberg}}, \bibinfo
  {author} {\bibfnamefont {H.}~\bibnamefont {Wang}},\ and\ \bibinfo {author}
  {\bibfnamefont {R.~J.}\ \bibnamefont {Wood}},\ }\bibfield  {title} {\bibinfo
  {title} {Controllable water surface to underwater transition through
  electrowetting in a hybrid terrestrial-aquatic microrobot},\ }\href@noop {}
  {\bibfield  {journal} {\bibinfo  {journal} {Nature Communications}\ }\textbf
  {\bibinfo {volume} {9}} (\bibinfo {year} {2018})}\BibitemShut {NoStop}%
\bibitem [{\citenamefont {Timm}\ \emph {et~al.}(2021)\citenamefont {Timm},
  \citenamefont {Kang}, \citenamefont {Rothstein},\ and\ \citenamefont
  {Masoud}}]{timm2021remotely}%
  \BibitemOpen
  \bibfield  {author} {\bibinfo {author} {\bibfnamefont {M.~L.}\ \bibnamefont
  {Timm}}, \bibinfo {author} {\bibfnamefont {S.~J.}\ \bibnamefont {Kang}},
  \bibinfo {author} {\bibfnamefont {J.~P.}\ \bibnamefont {Rothstein}},\ and\
  \bibinfo {author} {\bibfnamefont {H.}~\bibnamefont {Masoud}},\ }\bibfield
  {title} {\bibinfo {title} {A remotely controlled marangoni surfer},\
  }\href@noop {} {\bibfield  {journal} {\bibinfo  {journal} {Bioinspiration \&
  Biomimetics}\ }\textbf {\bibinfo {volume} {16}} (\bibinfo {year}
  {2021})}\BibitemShut {NoStop}%
\bibitem [{\citenamefont {Rhee}\ \emph {et~al.}(2022)\citenamefont {Rhee},
  \citenamefont {Hunt}, \citenamefont {Thomson},\ and\ \citenamefont
  {Harris}}]{rhee2022surferbot}%
  \BibitemOpen
  \bibfield  {author} {\bibinfo {author} {\bibfnamefont {E.}~\bibnamefont
  {Rhee}}, \bibinfo {author} {\bibfnamefont {R.}~\bibnamefont {Hunt}}, \bibinfo
  {author} {\bibfnamefont {S.~J.}\ \bibnamefont {Thomson}},\ and\ \bibinfo
  {author} {\bibfnamefont {D.~M.}\ \bibnamefont {Harris}},\ }\bibfield  {title}
  {\bibinfo {title} {Surferbot: a wave-propelled aquatic vibrobot},\
  }\href@noop {} {\bibfield  {journal} {\bibinfo  {journal} {Bioinspiration \&
  Biomimetics}\ }\textbf {\bibinfo {volume} {17}} (\bibinfo {year}
  {2022})}\BibitemShut {NoStop}%
\bibitem [{\citenamefont {Bush}\ and\ \citenamefont
  {Hu}(2006)}]{bush2006walking}%
  \BibitemOpen
  \bibfield  {author} {\bibinfo {author} {\bibfnamefont {J.~W.}\ \bibnamefont
  {Bush}}\ and\ \bibinfo {author} {\bibfnamefont {D.~L.}\ \bibnamefont {Hu}},\
  }\bibfield  {title} {\bibinfo {title} {Walking on water: biolocomotion at the
  interface},\ }\href@noop {} {\bibfield  {journal} {\bibinfo  {journal}
  {Annual Review of Fluid Mechanics}\ }\textbf {\bibinfo {volume} {38}}
  (\bibinfo {year} {2006})}\BibitemShut {NoStop}%
\bibitem [{\citenamefont {Speirs}\ \emph {et~al.}(2023)\citenamefont {Speirs},
  \citenamefont {Belden},\ and\ \citenamefont {Hellum}}]{speirs2023capture}%
  \BibitemOpen
  \bibfield  {author} {\bibinfo {author} {\bibfnamefont {N.~B.}\ \bibnamefont
  {Speirs}}, \bibinfo {author} {\bibfnamefont {J.~L.}\ \bibnamefont {Belden}},\
  and\ \bibinfo {author} {\bibfnamefont {A.~M.}\ \bibnamefont {Hellum}},\
  }\bibfield  {title} {\bibinfo {title} {The capture of airborne particulates
  by rain},\ }\href@noop {} {\bibfield  {journal} {\bibinfo  {journal} {Journal
  of Fluid Mechanics}\ }\textbf {\bibinfo {volume} {958}} (\bibinfo {year}
  {2023})}\BibitemShut {NoStop}%
\bibitem [{\citenamefont {Lamb}\ \emph {et~al.}(2017)\citenamefont {Lamb},
  \citenamefont {Brun},\ and\ \citenamefont {Fuller}}]{lamb2017direct}%
  \BibitemOpen
  \bibfield  {author} {\bibinfo {author} {\bibfnamefont {M.~P.}\ \bibnamefont
  {Lamb}}, \bibinfo {author} {\bibfnamefont {F.}~\bibnamefont {Brun}},\ and\
  \bibinfo {author} {\bibfnamefont {B.~M.}\ \bibnamefont {Fuller}},\ }\bibfield
   {title} {\bibinfo {title} {Direct measurements of lift and drag on shallowly
  submerged cobbles in steep streams: Implications for flow resistance and
  sediment transport},\ }\href@noop {} {\bibfield  {journal} {\bibinfo
  {journal} {Water Resources Research}\ }\textbf {\bibinfo {volume} {53}}
  (\bibinfo {year} {2017})}\BibitemShut {NoStop}%
\bibitem [{\citenamefont {Singh}\ \emph {et~al.}(2006)\citenamefont {Singh},
  \citenamefont {Nir},\ and\ \citenamefont {Semiat}}]{singh2006free}%
  \BibitemOpen
  \bibfield  {author} {\bibinfo {author} {\bibfnamefont {A.}~\bibnamefont
  {Singh}}, \bibinfo {author} {\bibfnamefont {A.}~\bibnamefont {Nir}},\ and\
  \bibinfo {author} {\bibfnamefont {R.}~\bibnamefont {Semiat}},\ }\bibfield
  {title} {\bibinfo {title} {Free-surface flow of concentrated suspensions},\
  }\href@noop {} {\bibfield  {journal} {\bibinfo  {journal} {International
  Journal of Multiphase Flow}\ }\textbf {\bibinfo {volume} {32}} (\bibinfo
  {year} {2006})}\BibitemShut {NoStop}%
\bibitem [{\citenamefont {Zhou}\ \emph {et~al.}(2022)\citenamefont {Zhou},
  \citenamefont {Vlahovska},\ and\ \citenamefont {Miksis}}]{zhou2022drag}%
  \BibitemOpen
  \bibfield  {author} {\bibinfo {author} {\bibfnamefont {Z.}~\bibnamefont
  {Zhou}}, \bibinfo {author} {\bibfnamefont {P.~M.}\ \bibnamefont
  {Vlahovska}},\ and\ \bibinfo {author} {\bibfnamefont {M.~J.}\ \bibnamefont
  {Miksis}},\ }\bibfield  {title} {\bibinfo {title} {Drag force on spherical
  particles trapped at a liquid interface},\ }\href@noop {} {\bibfield
  {journal} {\bibinfo  {journal} {Physical Review Fluids}\ }\textbf {\bibinfo
  {volume} {7}} (\bibinfo {year} {2022})}\BibitemShut {NoStop}%
\bibitem [{\citenamefont {Pham}\ \emph {et~al.}(2005)\citenamefont {Pham},
  \citenamefont {Nore},\ and\ \citenamefont {Brachet}}]{pham2005critical}%
  \BibitemOpen
  \bibfield  {author} {\bibinfo {author} {\bibfnamefont {C.-T.}\ \bibnamefont
  {Pham}}, \bibinfo {author} {\bibfnamefont {C.}~\bibnamefont {Nore}},\ and\
  \bibinfo {author} {\bibfnamefont {M.-{\'E}.}\ \bibnamefont {Brachet}},\
  }\bibfield  {title} {\bibinfo {title} {Critical speed for capillary-gravity
  surface flows in the dispersive shallow water limit},\ }\href@noop {}
  {\bibfield  {journal} {\bibinfo  {journal} {Physics of Fluids}\ }\textbf
  {\bibinfo {volume} {17}} (\bibinfo {year} {2005})}\BibitemShut {NoStop}%
\bibitem [{\citenamefont {Benzaquen}\ \emph {et~al.}(2011)\citenamefont
  {Benzaquen}, \citenamefont {Chevy},\ and\ \citenamefont
  {Rapha{\"e}l}}]{benzaquen2011wave}%
  \BibitemOpen
  \bibfield  {author} {\bibinfo {author} {\bibfnamefont {M.}~\bibnamefont
  {Benzaquen}}, \bibinfo {author} {\bibfnamefont {F.}~\bibnamefont {Chevy}},\
  and\ \bibinfo {author} {\bibfnamefont {{\'E}.}~\bibnamefont {Rapha{\"e}l}},\
  }\bibfield  {title} {\bibinfo {title} {Wave resistance for capillary gravity
  waves: finite-size effects},\ }\href@noop {} {\bibfield  {journal} {\bibinfo
  {journal} {Europhysics Letters}\ }\textbf {\bibinfo {volume} {96}} (\bibinfo
  {year} {2011})}\BibitemShut {NoStop}%
\end{thebibliography}%

\end{document}